\documentclass{article}

\usepackage[accepted]{icml2024}

\usepackage[backref=page]{hyperref}
\usepackage{amsmath}
\usepackage{amsfonts}
\usepackage{caption}
\usepackage{subcaption}
\usepackage{xcolor}
\usepackage{algorithm}
\usepackage{graphicx}

\icmltitlerunning{
    The Danger Of Arrogance: Welfare Equilibra As A Solution To Stackelberg Self-Play In Non-Coincidental Games
}

\begin{document}

\twocolumn[
\icmltitle{
    The Danger Of Arrogance: Welfare Equilibra As A Solution To \\
    Stackelberg Self-Play In Non-Coincidental Games
}



\icmlsetsymbol{equal}{*}

\begin{icmlauthorlist}
\icmlauthor{Jake Levi}{oxcs}
\icmlauthor{Chris Lu}{oxeng}
\icmlauthor{Timon Willi}{oxeng}
\icmlauthor{Christian Schroeder de Witt}{oxeng}
\icmlauthor{Jakob Foerster}{oxeng}
\end{icmlauthorlist}

\icmlaffiliation{oxcs}{Department of Computer Science, University of Oxford, Oxford, United Kingdom}
\icmlaffiliation{oxeng}{Department of Engineering, University of Oxford, Oxford, United Kingdom}

\icmlcorrespondingauthor{Jake Levi}{jake.levi@stcatz.ox.ac.uk}

\icmlkeywords{Multi-Agent Learning, Game Theory, General-Sum Games, Stackelberg, Opponent Shaping}

\vskip 0.3in
]



\printAffiliationsAndNotice{}  

\begin{abstract}
    The increasing prevalence of multi-agent learning systems in society necessitates understanding how to learn effective and safe policies in general-sum multi-agent environments against a variety of opponents, including self-play. General-sum learning is difficult because of non-stationary opponents and misaligned incentives. Our first main contribution is to show that many recent approaches to general-sum learning can be derived as approximations to \textit{Stackelberg strategies}, which suggests a framework for developing new multi-agent learning algorithms. We then define \textit{non-coincidental} games as games in which the Stackelberg strategy profile is not a Nash Equilibrium. This notably includes several canonical matrix games and provides a normative theory for why existing algorithms fail in self-play in such games. We address this problem by introducing Welfare Equilibria (WE) as a generalisation of Stackelberg Strategies, which can recover desirable Nash Equilibria even in non-coincidental games. Finally, we introduce Welfare Function Search (WelFuSe) as a practical approach to finding desirable WE against unknown opponents, which finds more mutually desirable solutions in self-play, while preserving performance against naive learning opponents.
\end{abstract}

\section{Introduction}
\label{section:intro}
The use of machine learning in multi-agent systems is becoming increasingly prevalent in society \cite{zhang2021multi}, but learning in multi-agent systems presents several challenges. Firstly, different agents might have misaligned or conflicting incentives \cite{subramanian2023robustness}, in which case agents won't necessarily act in the interests of other agents. Secondly, the learning environment is non-stationary, because other agents are simultaneously learning and updating their policies. Non-stationarity violates standard theoretical assumptions made in reinforcement learning \cite{sutton2018reinforcement}, and empirically makes learning significantly more challenging \cite{wang2022distributed}.

Real-world case-studies of multi-agent systems have shown how nefarious human agents can derail the behaviour of artificial learning agents, causing emotional harm to innocent observers and reputational damage to the learning agent's creators \cite{wolf2017we}. Such examples highlight the possible negative real-world consequences of deploying naively designed learning agents in multi-agent systems, and hence the necessity of adapting to the possible incentives and behaviour of other agents. These considerations are especially important in safety-critical environments \cite{kiran2021deep}.

Opponent shaping (OS) is an approach to general-sum learning in which agents explicitly consider the opponent's incentives and behaviour, and adapt their own behaviour in order to \textit{shape} the opponent's future learning process. For example, under self-play LOLA \cite{foerster2018learning} converges to prosocial solutions which incentivise opponent cooperation in canonical games such as the Iterated Prisoners' Dilemma (IPD). However, it has been shown \cite{willi2022cola} that OS algorithms can behave arrogantly in games such as the chicken game, leading to catastrophic outcomes in self-play. Avoiding catastrophe in self-play is fundamentally important for any learning algorithm that is to be deployed outside of simulation, because otherwise a malicious opponent could derail performance simply by choosing to use the same learning algorithm.

Our first contribution, in Section \ref{section:os stackelberg/solution concept}, is to show that Stackelberg strategies \cite{simaan1973stackelberg} when chosen by \emph{both} players represent a sensible solution concept in many two-player games. In Section \ref{section:os stackelberg/approximation} we show that many OS algorithms can be derived as approximations to Stackelberg strategies, and in Section \ref{section:os stackelberg/new os algorithms} we use this framework to derive new example algorithms which have qualitative advantages over existing approaches in small games. In Section \ref{section:we/nc games} we introduce \textit{non-coincidental games} as games in which the Stackelberg strategy profile is not a Nash Equilibrium, which notably includes several canonical matrix games (such as the chicken game), and helps explain why several OS algorithms which approximate Stackelberg strategies also fail in similar cases. We address this problem in Section \ref{section:we/chicken catastrophe} by introducing Welfare Equilibria (WE) as an abstract generalisation of Stackelberg strategies, which can find \textit{desirable} solutions in self-play in a broader class of games, including non-coincidental games. Lastly, in Section \ref{section:we/welfuse} we introduce Welfare Function Search (WelFuSe) as a practical approach to finding desirable WE against unknown opponents, which considers the problem of choosing a welfare function as a bandit problem, which is solved using posterior sampling. We demonstrate how WelFuSe finds more desirable solutions in OS self-play while preserving performance against naive learning opponents. The payoff tables for all matrix games we describe are included in Table \ref{table:matrix game payoff tables} in the Appendix, and code is publicly available online\footnote{\href{https://github.com/jakelevi1996/welfare_equilibria_public}{{https://github.com/jakelevi1996/welfare\_equilibria\_public}}}.

\section{Background}
\label{section:related work}
Agents in multi-agent systems may have misaligned or conflicting incentives, which can be analysed using game theory. A game consists of a fixed number of players, each of which chooses a strategy, and the collection of strategies chosen by each player is known as the strategy profile. Each player has a reward function, which returns to them a scalar reward as a function of any given strategy profile. A given player's strategy is a best response (BR) to a given strategy profile when there is no other strategy available to that player which increases the value of their reward function (assuming the rest of the strategy profile remains fixed). A strategy profile is a Nash Equilibrium (NE) when every player's strategy is a BR to all other players, in which case no agent is incentivised to deviate from their choice of strategy \cite{nash1950equilibrium}.

A \textbf{Stackelberg strategy} is an abstract choice of strategy that either player can choose to play in almost any two-player game \cite{simaan1973stackelberg}, modelled on the concept of Stackelberg games. Suppose a two-player game $G$ consists of strategies $x$ and $y$ (one for each player) and reward functions $R^x(x, y)$ and $R^y(x, y)$, and $x$ is chosen to be a Stackelberg strategy. The player that chooses $x$ assumes the opponent will be able to first observe and then play a BR to any given choice of $x$, therefore the opponent's strategy can be described by the ``Opponent BR \emph{Function}'', denoted by $y^*(x)$ and shown below in Equation \ref{eq:y BR function}. Once the opponent BR function is known, the reward for any given choice of $x$ is a function only of $x$ (and not of $y$), which can be maximised using standard optimisation approaches, leading to the Stackelberg strategy $x^*$, shown in Equation \ref{eq:stackelberg x}. The opponent BR functions and Stackelberg strategies for either player in the game $G$ are summarised below:

\begin{align}
    x^* (y) &= \underset{x}{\mathrm{argmax}}\Bigl[ R^x(x, y) \Bigr] \label{eq:x BR function} \\
    y^* (x) &= \underset{y}{\mathrm{argmax}}\Bigl[ R^y(x, y) \Bigr] \label{eq:y BR function} \\
    x^* &= \underset{x}{\mathrm{argmax}}\left[R^x\Bigl(x, y^* (x) \Bigr)\right] \label{eq:stackelberg x} \\
    y^* &= \underset{y}{\mathrm{argmax}}\left[R^y\Bigl(x^* (y), y \Bigr)\right] \label{eq:stackelberg y}
\end{align}

A simple approach to \emph{learning} in multi-agent systems is \textit{naive learning} (NL). A NL agent takes steps of gradient ascent on their own reward function, given the current strategies of all other players at each time step, assuming those strategies are constant. There are many simple scenarios in which NL fails to converge in self-play \cite{singh2000nash}. OS approaches address this problem by anticipating and shaping the update step of the opponent. The first OS method is LOLA \cite{foerster2018learning}, which includes a Taylor series expansion of its reward function after simulating an opponent NL update in its optimisation objective. LOLA achieved impressive results such as converging to the tit-for-tat (TFT) NE strategy in IPD self-play. However, later work \cite{letcher2018stable} showed that LOLA can fail to preserve NE, and introduced SOS as an alternative, which interpolates between LOLA and an earlier approach known as LookAhead \cite{zhang2010multi}. This addresses the problem of ``arrogance'' and encourages SOS to converge to stable-fix points. COLA \cite{willi2022cola} addresses the inconsistency in which LOLA implicitly assumes that the opponent is a NL agent by explicitly learning the opponent's update function, and using the learned update function to train a policy. The same work also introduced Exact-LOLA (ELOLA), which is similar to LOLA except it does not approximate the perturbed reward function using Taylor series. M-FOS \cite{lu2022model,fung2023analyzing} achieves particularly good performance against NL agents by learning which strategy to play next as a function of both agents' most recently played strategies. This function does not use an explicit model of the opponent, and is learned over multiple episodes against the opponent. SHAPER \cite{khan2023scaling} is parameterised with an RNN which captures memory over multiple time scales, allowing it to scale up to high-dimensional and $n$-player games \cite{souly2023leading}. The Good Shepherd \cite{balaguer2022good} anticipates the opponent's long-term behaviour by simulating many naive updates of the opponent's parameters, and using these hypothetical future opponent parameters when optimising the reward function.

OS approaches have found desirable solutions in a variety of classic matrix games, as well as extensive-form games such as the Coin Game \cite{foerster2018learning}. However, there are differentiable games which do not contain any NE, in which many approaches (including OS approaches) fail to converge. Even defining a solution concept in such games, which do not contain any NE, is non-trivial. In Section \ref{section:os stackelberg/solution concept} we focus on one such example (provided by \cite{letcher2020impossibility} in the proof of Theorem 1, page 5) which we refer to as the ``Impossible Market'', which consists of two players choosing strategies $x \in \mathbb{R}$ and $y \in \mathbb{R}$ with the following reward functions:

\begin{align}
    R^x(x, y) &= - \frac{x^6}{6} + \frac{x^2}{2} - xy - \frac{1}{4}\left( \frac{y^4}{1 + x^2} - \frac{x^4}{1 + y^2} \right) \\
    R^y(x, y) &= - \frac{y^6}{6} + \frac{y^2}{2} + xy + \frac{1}{4}\left( \frac{y^4}{1 + x^2} - \frac{x^4}{1 + y^2} \right)
\end{align}

\section{Related Work}

Much previous work has explored Stackelberg strategies in various applications \cite{pita2009using, yin2010stackelberg, yin2012trusts}, and has also argued that Stackelberg equilibria (in which a leader plays a Stackelberg strategy and a follower plays a BR) should be considered as solution concepts, even in games which are not Stackelberg games \cite{conitzer2016stackelberg}. Using Query Oracles as an abstraction for the opponent BR has facilitated the extension of Stackelberg strategies to deep reinforcement learning \cite{gerstgrasser2023oracles}. Using commitment schedules has allowed Stackelberg strategies to be learned without access to the opponent reward function \cite{loftin2023uncoupled}. However, a common theme in previous work on Stackelberg strategies is the assumption that the opponent always plays a BR to the Stackelberg strategy, introducing possibly arbitrary asymmetry into the game, rather than considering the implications of both players choosing Stackelberg strategies. In contrast, we will consider Stackelberg strategies in self-play, and generalisations of Stackelberg strategies which display more desirable behaviour in self-play in a broader range of environments.

Beyond Stackelberg strategies, previous work has investigated alternative solution concepts and approaches to learning and equilibrium selection. An Active Markov Game accounts for agents in a game updating their strategies over time, and an Active Equilibrium is a solution concept in which no agent could improve their long term average reward by changing their update function \cite{kim2022influencing, kim2022game}. FURTHER is an approach for learning in an Active Markov Game using variational inference to approximate opponent update functions \cite{kim2022influencing}. The Active Equilibrium solution concept is closely related to a NE in a meta-game, in which the strategy in the meta-game is the update function, and the reward in the meta-game is long-term average reward over time. This interpretation illuminates the connection with MFOS \cite{lu2022model}, which uses reinforcement learning to shape the opponent learning in the meta-game. Modelling Opponent Learning (MOL) uses a two-phase approach, which first learns the game structure and BR of the opponent, and then guides the opponent's learning in the second phase \cite{hu2023modeling}.

\section{Problem Settings}

We consider two different problem settings. The first, in Section \ref{section:os stackelberg}, is a traditional general-sum learning setting similar to that used by LOLA \cite{foerster2018learning}, in which both agents have full access to the environment and the opponent's most recent strategy, including gradients. This could apply to
an offline learning setting against a simulated opponent, after which the parameters are frozen and deployed online.
The second problem setting in Section \ref{section:we} assumes a \textit{meta-learning} approach, which uses batches and multiple episodes of learning.
Agent strategies are reset between each episode, which has more in common with general-sum meta-learning approaches such as MFOS \cite{lu2022model}. The resulting \textit{shaping strategy} is then deployed in the real world.
This assumes that at test time the \textit{shaper} interacts with the \textit{shapee} over the course of the shapee's training horizon.

\section{The Stackelberg Framework For General-Sum Learning}
\label{section:os stackelberg}
\subsection{The Stackelberg Strategy Profile As A Solution Concept} \label{section:os stackelberg/solution concept}
Stackelberg strategies are usually motivated by assuming asymmetry between players, such as one player learning more quickly than the other \cite{simaan1973stackelberg} or having extra information \cite{chen1972stackelburg}. However, there are situations in which it is sensible for both players to bilaterally choose Stackelberg strategies, such as in the Impossible Market which was introduced in Section \ref{section:related work}, which does not contain any NE. It is straightforward to approximate the opponent BR functions and Stackelberg strategies for each player in the Impossible Market using grid search. The results of doing so are shown in Figure \ref{fig:ImpossibleMarket_greedy} in the Appendix, leading to the Stackelberg strategies $x^* = 0$ and $y^* = 0$. If $x$ and $y$ both play Stackelberg strategies in the Impossible Market, they both receive a reward of 0. Even if either player switches to a BR, the Stackelberg agent receives a reward of -0.654. This outcome is still better for the Stackelberg agent than the worst-case reward that both agents would periodically receive if they both used learning algorithms which converge to a limit cycle \cite{letcher2020impossibility}, which is typically less than -1. Furthermore, both players cannot play BR strategies simultaneously because the game does not contain any NE. The Stackelberg strategy therefore represents a robust choice of strategy for each agent in the Impossible Market.

In general it is possible for both agents to choose Stackelberg strategies in any two-player game, and therefore the Stackelberg strategy profile (the strategy profile in which \emph{both} players choose Stackelberg strategies) represents a solution concept for two-player games. Notably, the Stackelberg strategy profile may be distinct from a Stackelberg equilibrium, in which one player plays a Stackelberg strategy and the opponent plays a BR \cite{conitzer2016stackelberg}. Unlike NE, the Stackelberg strategy profile provides a unique solution concept in almost every two-player game (whereas games may have multiple NE), and the Stackelberg strategy profile also generalises to two-player games which do not contain any NE (such as the Impossible Market). Included in the Appendix are depictions of the Stackelberg strategy profile for the games of Matching Pennies (Figure \ref{fig:MatchingPennies_greedy}), Stag Hunt (Figure \ref{fig:StagHunt_greedy}), Prisoners' Dilemma (Figure \ref{fig:PrisonersDilemma_greedy}), Awkward Game (an asymmetric, general-sum, two-player $2 \times 2$ matrix game, containing only one NE which is mixed, shown in Figure \ref{fig:AwkwardGame_greedy}), and \texttt{IpdTftAlldMix} (a version of IPD in which both players must choose a 1D strategy, referring to a parameter which interpolates between the all-defect and TFT strategies, shown in Figure \ref{fig:IpdTftAlldMix_greedy}), showing that the Stackelberg strategy profile locates the most desirable NE (in terms of maximising both players' rewards) in all five cases, regardless of whether that NE is pure or mixed.

\subsection{Opponent-Shaping Derived Via The Stackelberg Framework} \label{section:os stackelberg/approximation}

It is straightforward to approximate the BR functions and Stackelberg strategies in the Impossible Market with grid search because the strategies for both players are 1D. In games with high-dimensional action spaces it will be impractical to use grid search, and other types of approximation are needed. Here, we demonstrate that approximating $\mathrm{argmax}$ in several different ways recovers different OS algorithms, as well as  LookAhead. LookAhead is an ``opponent-aware'' algorithm which, while aware of the opponent's update step, does not shape it.

Suppose that at time $t$ our strategy is $x_t$ and the opponent's strategy is $y_t$. If we approximate $\hat{y}_t \approx y^*(x_t)$ in Equation \ref{eq:stackelberg x} with a single step of gradient ascent with learning rate $\alpha$, and we approximate $x_{t+1} \approx x^*$ with a single step of gradient ascent with learning rate $\eta$ (without gradient flow from $\hat{y}_t$ to $x_{t+1}$), we recover the LookAhead learning update:

\begin{equation}
    \begin{aligned}
        x_{t+1} &= x_t + \eta \frac{\partial}{\partial x_t}\left[R^x\Bigl(x_t, \hat{y_t} \Bigr)\right] \\
        \mathrm{where} \quad \hat{y_t} &= y_t + \alpha \frac{\partial}{\partial y_t}\Bigl[R^y\left(x_t, y_t \right)\Bigr]
    \end{aligned}
\end{equation}

If instead we do include gradient flow from $\hat{y}_t(x_t)$ (now written as a function of $x_t$) to $x_{t+1}$, we recover the ELOLA learning update:

\begin{equation}
    \begin{aligned}
        x_{t+1} &= x_t + \eta \frac{\partial}{\partial x_t}\left[R^x\Bigl(x_t, \hat{y_t}(x_t) \Bigr)\right] \\
        \mathrm{where} \quad \hat{y_t}(x_t) &= y_t + \alpha \frac{\partial}{\partial y_t}\Bigl[R^y\left(x_t, y_t \right)\Bigr]
    \end{aligned}
\end{equation}

If we also approximate the reward function $R^x$ with a first-order Taylor series with respect to $\hat{y}_t(x_t)$ then we recover the original LOLA update. If we interpolate between LOLA and LookAhead (which have both been shown to be approximations of Stackelberg strategies) then we recover SOS (which is therefore also an approximation of a Stackelberg strategy). If we approximate $y^*(x)$ in Equation \ref{eq:stackelberg x} with many steps of gradient ascent (while still approximating $x^*$ with a single step of gradient ascent) then we recover the Good Shepherd learning update.

The OS algorithms we considered so far use explicit update rules for choosing strategies on successive steps of a game. MFOS is distinct from such algorithms in the sense that MFOS learns an update rule which is fixed between time steps but varied between episodes, in order to maximise the expected discounted return in each episode. The MFOS update rule is learned from experience against its opponent, and therefore the extent to which MFOS fits in to the Stackelberg framework depends on the nature of the opponent. Against an opponent which always plays an approximate BR (which is a reasonable approximation for a rational opponent), MFOS learns to play the strategy which maximises reward against the opponent stategy, which approximates the Stackelberg strategy to the extent that the opponent plays an approximate BR. This interpretation is consistent with the observation that MFOS learns to play a ZD extortion strategy against a ``Look-Ahead Best Response'' agent in IPD \cite{lu2022model}, which is the best strategy to play against an opponent that plays an approximate BR, and is therefore the Stackelberg strategy in IPD.

\subsection{Deriving New Approximate Stackelberg Learning Algorithms} \label{section:os stackelberg/new os algorithms}

\begin{figure*}
    \centering
    \captionsetup[subfigure]{justification=centering}
    \begin{subfigure}{0.3\textwidth}
        \centering
        \includegraphics[width=\textwidth]{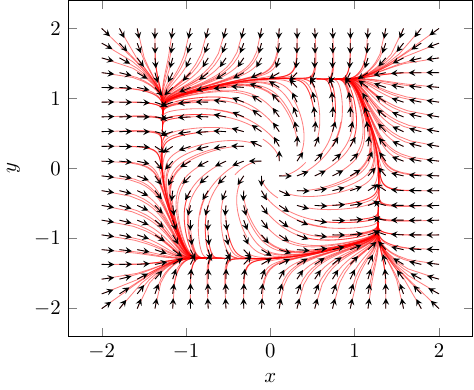}
        \caption{NL self-play \\ $\eta=0.01$}
        \label{fig:ImpossibleMarket NL self-play}
    \end{subfigure}
    \hfill
    \begin{subfigure}{0.3\textwidth}
        \centering
        \includegraphics[width=\textwidth]{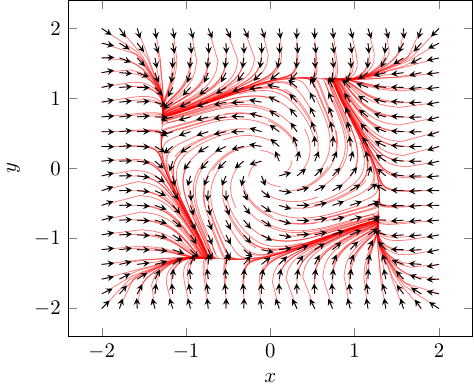}
        \caption{ELOLA self-play \\ $\eta=0.01, \alpha=0.2$}
        \label{fig:ImpossibleMarket ELOLA self-play}
    \end{subfigure}
    \hfill
    \begin{subfigure}{0.3\textwidth}
        \centering
        \includegraphics[width=\textwidth]{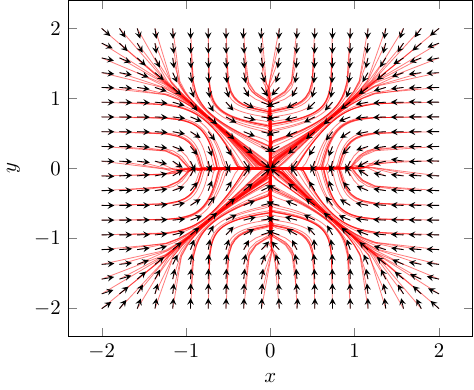}
        \caption{SaGa self-play \\ $\eta=0.01, N=200, \sigma=1$}
        \label{fig:ImpossibleMarket SaGa self-play}
    \end{subfigure}
    \caption{Self-play phase portraits for different algorithms in the Impossible Market. Gradient arrow-lengths are normalised for visual clarity and do not represent gradient magnitude.}
    \label{fig:ImpossibleMarket self-play}
\end{figure*}

In Section \ref{section:os stackelberg/approximation} we showed that several OS algorithms approximate Stackelberg strategies. In this subsection we demonstrate that different approximations of the Stackelberg strategy can be used to devise new learning algorithms with qualitative advantages over existing approaches. Our intention here is simply to demonstrate the possibilities available within this framework. Developing these specific algorithms and further exploring the possibilities within this framework will be explored in future work.

In Section \ref{section:os stackelberg/solution concept} we demonstrated that Stackelberg strategies offer a sensible solution concept in the Impossible Market. However previous work \cite{letcher2020impossibility} has shown that many algorithms (including several OS algorithms) actually fail to converge in the Impossible Market. Suppose we approximate the inner $\mathrm{argmax}$ used to calculate $y^*(x_t)$ in Equation \ref{eq:stackelberg x} by randomly sampling $N$ different opponent strategies, calculating the opponent's reward for each one, and assuming the opponent chooses the strategy which maximises their reward. We then update our own strategy using a single step of gradient ascent on our own reward function, evaluated using our current strategy and this approximate opponent BR. This can be summarised in the following learning update:

\begin{equation}
    \begin{aligned}
        x_{t+1} &= x_t + \eta \frac{\partial}{\partial x_t}\Bigl[R^x\left(x_t, \hat{y_t} \right)\Bigr] \\
        \mathrm{where} \quad &\begin{cases} \hat{y_t} = \underset{n \in \{1, \dots, N\}}{\mathrm{max}}\Bigl[ R^y(x_t, y_t + \varepsilon_n) \Bigr] \\ \varepsilon_n \sim \mathcal{N}(0, \sigma^2) \end{cases}
    \end{aligned}
\end{equation}

This update uses a SAmpling-based approximation to the inner $\mathrm{argmax}$ and a Gradient-Ascent-based approximation to the outer $\mathrm{argmax}$, so we refer to this learning algorithm as ``SaGa''. The behaviour of SaGa self-play in the Impossible Market is shown in a phase portrait in Figure \ref{fig:ImpossibleMarket self-play}, alongside equivalent phase portraits for NL and ELOLA. These results demonstrate how NL and ELOLA self-play both consistently converge to qualitatively similar limit-cycles, whereas SaGa self-play consistently converges to the Stackelberg strategy profile, simply by using a sampling-based approximation to the opponent BR.

Like any algorithm which uses gradient ascent, existing OS algorithms are vulnerable to getting stuck in non-global local maxima. This problem is particularly prevalent for games such as Stag Hunt and \texttt{IpdTftAlldMix}, for which the reward function against a perfect BR opponent is not concave (see Figures \ref{fig:StagHunt_greedy} and \ref{fig:IpdTftAlldMix_greedy} in the Appendix). It is therefore natural to consider whether we can achieve more robust performance by using sampling-based approximations for the outer $\mathrm{argmax}$ as well as the inner $\mathrm{argmax}$, which can be achieved using the following learning update:

\begin{equation}
    \begin{aligned}
        x_{t+1} &= (1-\eta) x_t + \eta \underset{m\in\{1,\dots,M\}}{\mathrm{max}}\left[R^x\Bigl(x_t + \varepsilon^x_m, \hat{y_t}\left(\varepsilon^x_m\right) \Bigr)\right] \\
        \mathrm{where}& \quad \begin{cases} \hat{y_t}\left(\varepsilon^x_m\right) &= \underset{n\in\{1,\dots,N\}}{\mathrm{max}}\Bigl[ R^y(x_t + \varepsilon^x_m, y_t + \varepsilon^y_n) \Bigr] \\ \varepsilon^y_n, \varepsilon^x_m &\sim \mathcal{N}(0, \sigma^2) \end{cases}
    \end{aligned}
\end{equation}

\begin{figure*}
    \centering
    \captionsetup[subfigure]{justification=centering}
    \begin{subfigure}{0.3\textwidth}
        \centering
        \includegraphics[width=\textwidth]{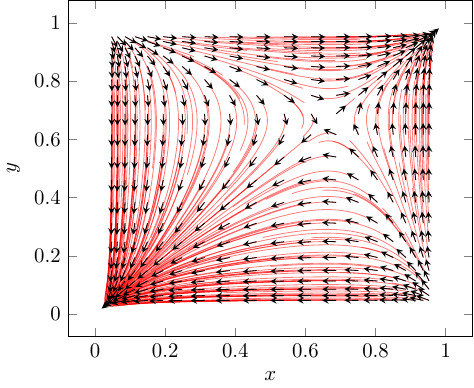}
        \caption{NL self-play \\ $\eta=0.1$}
        \label{fig:StagHunt NL self-play}
    \end{subfigure}
    \hfill
    \begin{subfigure}{0.3\textwidth}
        \centering
        \includegraphics[width=\textwidth]{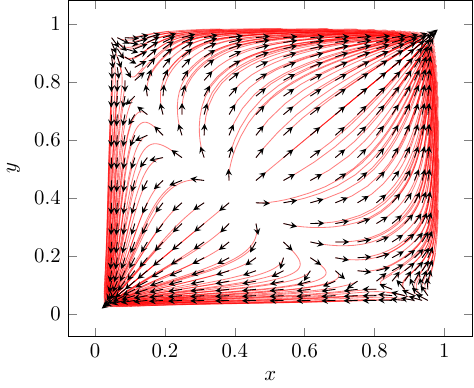}
        \caption{ELOLA self-play \\ $\eta=0.1, \alpha=5$}
        \label{fig:StagHunt ELOLA self-play}
    \end{subfigure}
    \hfill
    \begin{subfigure}{0.3\textwidth}
        \centering
        \includegraphics[width=\textwidth]{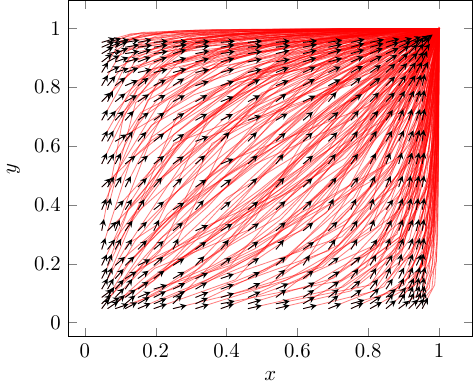}
        \caption{SaSa self-play \\ $\eta=0.1, M=N=50, \sigma=5$}
        \label{fig:StagHunt SaSa self-play}
    \end{subfigure}
    \caption{Self-play phase portraits for different algorithms in the Stag Hunt game. Axes refer to probability of hunting a stag for each player respectively, with the upper right corner corresponding to the optimal NE.}
    \label{fig:StagHunt self-play}
\end{figure*}

This update uses a SAmpling-based approximation to the inner $\mathrm{argmax}$ \emph{and} a SAmpling-based approximation to the outer $\mathrm{argmax}$, so we refer to this learning algorithm as ``SaSa''. The behaviour of SaSa self-play in the Stag Hunt game is shown in a phase portrait in Figure \ref{fig:StagHunt self-play}, alongside equivalent phase portraits for NL and ELOLA. These results demonstrate how NL and ELOLA self-play both converge to sub-optimal NE from relatively large areas of the state-space, whereas SaSa self-play consistently converges to the optimal NE, simply by using sampling-based $\mathrm{argmax}$ approximations.

\begin{figure*}
    \centering
    \begin{subfigure}{0.3\textwidth}
        \centering
        \includegraphics[width=\textwidth]{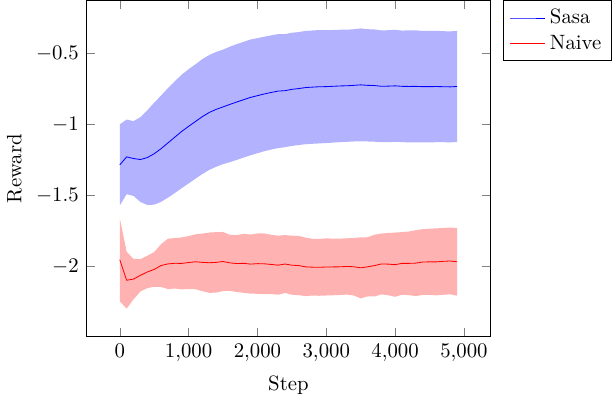}
        \caption{Mean rewards}
    \end{subfigure}
    \hfill
    \begin{subfigure}{0.3\textwidth}
        \centering
        \includegraphics[width=\textwidth]{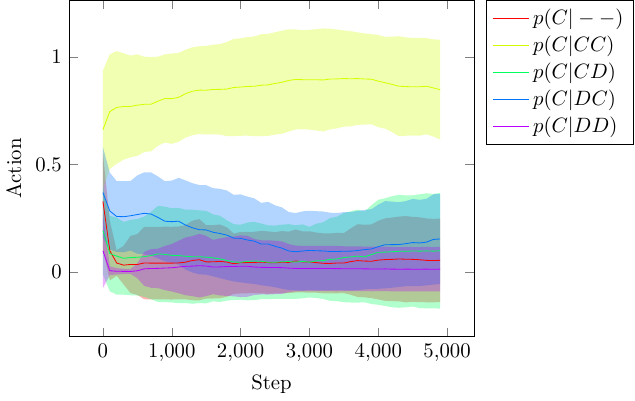}
        \caption{SaSa mean actions}
    \end{subfigure}
    \hfill
    \begin{subfigure}{0.3\textwidth}
        \centering
        \includegraphics[width=\textwidth]{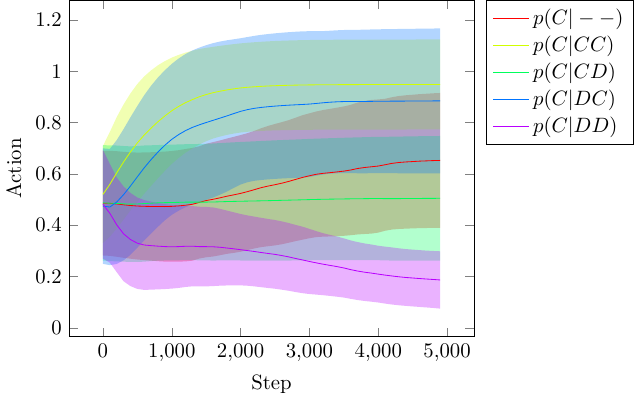}
        \caption{NL mean actions}
    \end{subfigure}
    \caption{Results for 5000 steps of learning with SaSa ($\eta=0.1,M=20,N=20,\sigma=1$) against NL ($\eta=0.1$) in IPD (using discounted returns with $\gamma = 0.96$), averaged over 100 trials. Final mean rewards are -0.72972 and -1.97547 for SaSa and NL respectively.}
    \label{fig:IPD saga vs NL}
\end{figure*}

The results for SaSa against NL in IPD are included in Figure \ref{fig:IPD saga vs NL}. SaSa achieves a stable mean reward of -0.730, which is greater than the reward for mutual-TFT and suggests that SaSa on average learns a ZD extortion strategy \cite{press2012iterated}, demonstrating that SaSa can scale up to games with moderately high-dimensional action spaces.

\section{Welfare Equilibria}
\label{section:we}
\subsection{Non-Coincidental Games And The Chicken Catastrophe} \label{section:we/nc games}
As mentioned in Section \ref{section:os stackelberg/solution concept}, Stackelberg strategy profiles find optimal NE in a variety of zero-sum and general-sum two-player games. A common property of many of these games is that the most rewarding NE for one player \emph{coincides} with the most rewarding NE for the other player, so we refer to such games as ``Coincidental games''. In general however, there are many games in which this property does not hold, such as the Chicken Game. The Stackelberg strategy for either player in the Chicken Game is to drive straight (to which the opponent BR is to chicken out), however if both players drive straight then they both experience the unique worst possible outcome, in which neither player's strategy is a BR (shown in Figure \ref{fig:ChickenGame_greedy} in the Appendix). We formally define non-coincidental games as games in which the Stackelberg strategy profile is not a NE\footnote{Interestingly, the Impossible Market is a non-coincidental game, although the Stackelberg strategy profile achieves desirable behaviour in this case. One response to this observation follows from the concept of arrogance penalties, which are introduced in Section \ref{section:we/arrogance penalty} of the Appendix. Specifically, the Impossible Market has negative arrogance penalties, so both players are rewarded for ``arrogantly'' assuming an opponent BR and choosing a Stackelberg strategy. This is in contrast to the other non-coincidental games we consider, wherein mutual arrogance has a cost.}, and refer to the failure of the Stackelberg strategy profile to find a desirable outcome in the chicken game (a non-coincidental canonical matrix game) as ``The Chicken Catastrophe''.

In Section \ref{section:os stackelberg/approximation} we showed that many OS algorithms approximate Stackelberg strategies. Considering that the Stackelberg strategy profile fundamentally fails in the Chicken Game, this provides a normative theory for why OS algorithms which approximate Stackelberg strategies \emph{should also} fail in self-play in the Chicken Game. This analysis is consistent with empirical evidence \cite{willi2022cola}, and is problematic due to the importance of finding desirable solutions in self-play (which we highlighted in Section \ref{section:intro}). In the following subsections we present a generalisation of Stackelberg strategies which is capable of finding NE in self-play in non-coincidental games. Our analysis excludes M-FOS meta-self-play and the SOS algorithm, which both address arrogance and are outside our framework.

\subsection{Addressing The Chicken Catastrophe} \label{section:we/chicken catastrophe}
A simple solution for the Chicken Catastrophe is for each player to maximise egalitarian social welfare (defined in Equation \ref{eq:egalitarian welfare}) instead of self-reward, while still assuming the opponent plays a BR. If both players take this approach, then they both concurrently select the unique mixed NE in the Chicken Game, and both receive a better reward than if they had both played Stackelberg strategies (shown in Figure \ref{fig:ChickenGame_egalitarian} in the Appendix). However, this approach fails in self-play in the non-coincidental game Bach Or Stravinsky (BOS). Both pure NE in BOS have maximal and equal egalitarian social welfare, so there is no way for both players to consistently favour one NE over the other in a way which is fair to both players. A solution to BOS self-play is for each player to maximise fairness (defined in Equation \ref{eq:fairness welfare}) while assuming the opponent plays a BR. If both players take this approach, then they both concurrently select the unique mixed NE in BOS, which is maximally fair to both players, and also better than failing to coordinate (shown in Figure \ref{fig:CoordinationGame_fairness}). However, this approach fails in self-play in a game we introduce and refer to as the Eagle Game. In the Eagle Game, maximising egalitarian social welfare leads to better outcomes for both players than maximising self-reward or fairness, assuming the opponent plays a BR in all cases (shown in Figure \ref{fig:EagleGame_egalitarian}).

The previous paragraph demonstrates that choosing an appropriate welfare function and then maximising that welfare function (while assuming that the opponent will play a \textit{greedy} BR) allows each player to effectively choose between desirable NE strategies in a wide variety of games. However, to our knowledge there is no single welfare function that consistently leads to desirable NE strategies in every possible game. This motivates us to define the ``Welfare Equilibrium'' (WE) strategy for \emph{any} given choice of welfare function as shown below, followed by examples of empircally useful welfare functions (the opponent BR functions are defined as in Equations \ref{eq:x BR function} and \ref{eq:y BR function}):

\begin{align}
    x^*_{\textrm{WE}} &= \underset{x}{\mathrm{argmax}}\left[w^x\Bigl(x, y^* (x) \Bigr)\right] \label{eq:we x} \\
    y^*_{\textrm{WE}} &= \underset{y}{\mathrm{argmax}}\left[w^y\Bigl(x^* (y), y \Bigr)\right] \label{eq:we y} \\
    w_{\textrm{greedy}}^x(x, y)     &= R^x(x, y) \\
    w_{\textrm{greedy}}^y(x, y)     &= R^y(x, y) \\
    w_{\textrm{egalitarian}}(x, y)  &= \min\Bigl( R^x(x, y), R^y(x, y) \Bigr) \label{eq:egalitarian welfare} \\
    w_{\textrm{fairness}}(x, y)     &= -\Bigl\vert R^x(x, y) - R^y(x, y) \Bigr\vert \label{eq:fairness welfare}
\end{align}

The Stackelberg strategy for either player is a special case of a WE strategy in which that player chooses to maximise a greedy welfare function (self-reward). The WE strategy is therefore a generalisation of the Stackelberg strategy. Included in the Appendix are depictions of WE strategy profiles for the Chicken Game (Figure \ref{fig:ChickenGame_egalitarian}), Baby Chicken Game (a variant of the Chicken Game with a less severe punishment for mutually driving straight, which leads to a more intelligible depiction of the WE profile, shown in Figure \ref{fig:BabyChickenGame_egalitarian}), BOS (Figure \ref{fig:CoordinationGame_fairness}), Tandem Game (Figure \ref{fig:Tandem_egalitarian}), Ultimatum Game (Figure \ref{fig:UltimatumGame_egalitarian}), and Eagle Game (Figure \ref{fig:EagleGame_egalitarian}), showing that WE strategy profiles with appropriate welfare functions can locate desirable NE and provide better rewards for both players than the equivalent Stackelberg strategy profile in all six cases.

A natural question to ask is when (if ever) it is actually in a player's interest to maximise any welfare function other than a greedy welfare function. The performance of any learning algorithm in a multi-agent system depends on the nature and dynamics of the opponent. Against an opponent that will always play a BR, the Stackelberg strategy (equivalent to a greedy WE strategy) is always the best strategy to play (by definition). However, as outlined in Section \ref{section:intro}, it is also important to consider how an algorithm performs in self-play, and for two WE agents in self-play in non-coincidental games, the greedy welfare function is often not the best welfare function to choose. This is demonstrated in Table \ref{table:we chicken}, which shows the possible rewards for WE agents in self-play choosing between greedy, egalitarian and fairness welfare functions and then playing WE strategies in the Tandem Game \cite{letcher2018stable}. In this example, not only is it sometimes in a player's interest to maximise a non-greedy welfare function, but in fact choosing the greedy welfare function is a \emph{strictly dominated strategy}. This means that for either player, regardless of whichever welfare function is chosen by the opponent, a better reward would be achieved by choosing egalitarian or fairness welfare functions instead of a greedy welfare function. Therefore in this example, both players have a \emph{greedy incentive to maximise a non-greedy welfare function}.

\begin{table}
    \caption{Rewards in Tandem Game for WE strategy profiles with different welfare functions. Strategies are restricted to the range $[-2, 3]$ for numerical tractability, as in Figure \ref{fig:Tandem_egalitarian}.}
    \label{table:we chicken}
    \begin{tabular}{l|c c c}
        \hline
        & \textcolor{violet}{Greedy (G)} & \textcolor{violet}{Egalitarian (E)} & \textcolor{violet}{Fairness (F)} \\
        \hline
        \textcolor{blue}{G} & (\textcolor{blue}{-30.00},    \textcolor{violet}{-30.00}) &   (\textcolor{blue}{-6.24}, \textcolor{violet}{-11.24})  &   (\textcolor{blue}{-6.24}, \textcolor{violet}{-11.24})  \\
        \textcolor{blue}{E} & (\textcolor{blue}{-11.24},    \textcolor{violet}{ -6.24}) &   (\textcolor{blue}{ 0.00}, \textcolor{violet}{  0.00})  &   (\textcolor{blue}{ 0.00}, \textcolor{violet}{  0.00})  \\
        \textcolor{blue}{F} & (\textcolor{blue}{-11.24},    \textcolor{violet}{ -6.24}) &   (\textcolor{blue}{ 0.00}, \textcolor{violet}{  0.00})  &   (\textcolor{blue}{ 0.00}, \textcolor{violet}{  0.00})  \\
        \hline
    \end{tabular}
\end{table}

The welfare functions considered so far are not invariant to affine transformations of either player's reward function. This can be addressed by introducing ``arrogance penalties'', which are discussed in Section \ref{section:we/arrogance penalty} of the Appendix.

\begin{figure*}
    \centering
    \begin{subfigure}{0.3\textwidth}
        \centering
        \includegraphics[width=\textwidth]{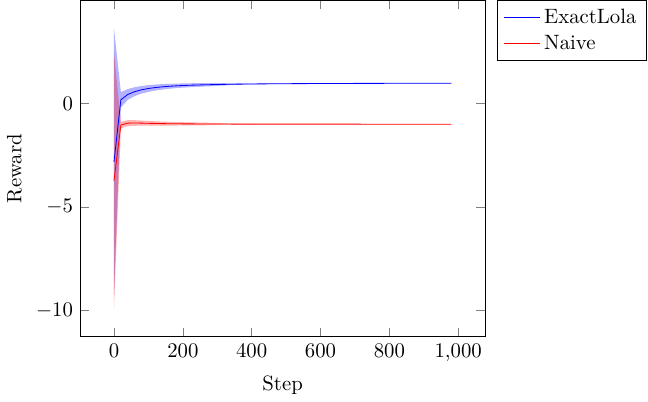}
        \caption{}
        \label{fig:elola nl chicken game}
    \end{subfigure}
    \hfill
    \begin{subfigure}{0.3\textwidth}
        \centering
        \includegraphics[width=\textwidth]{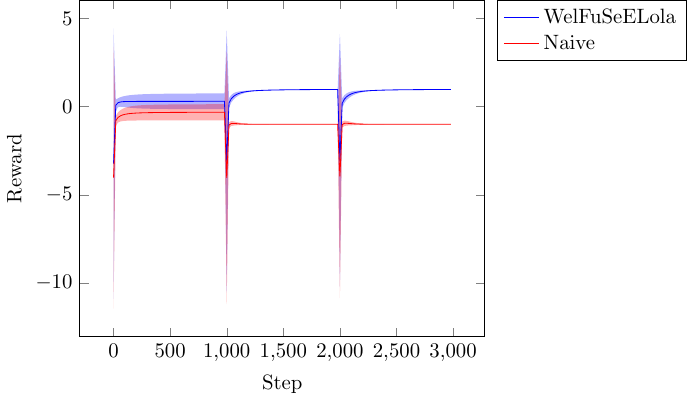}
        \caption{}
        \label{fig:welfuseelola nl chicken game rewards}
    \end{subfigure}
    \hfill
    \begin{subfigure}{0.3\textwidth}
        \centering
        \includegraphics[width=\textwidth]{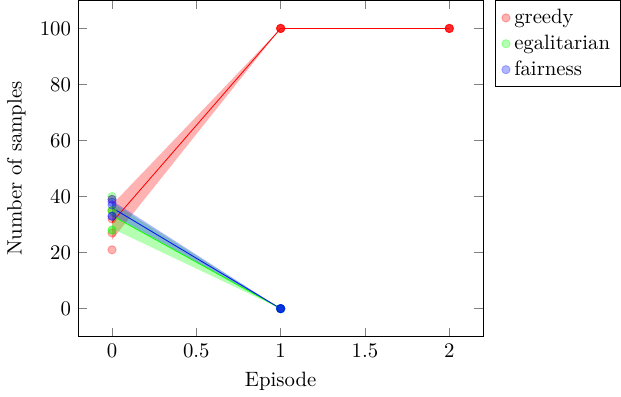}
        \caption{}
        \label{fig:welfuseelola nl chicken game welfare function history}
    \end{subfigure}
    \newline
    \begin{subfigure}{0.3\textwidth}
        \centering
        \includegraphics[width=\textwidth]{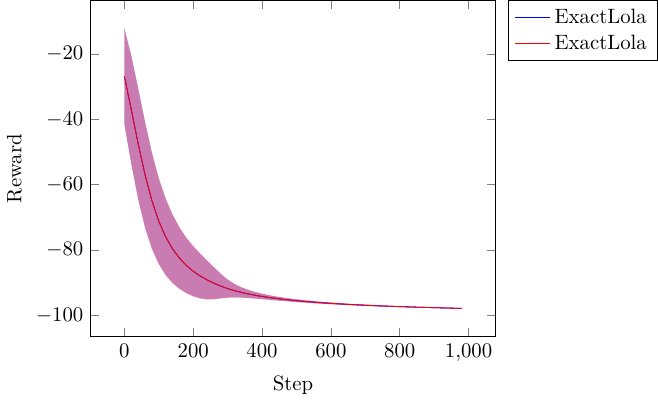}
        \caption{}
        \label{fig:elola sp chicken game}
    \end{subfigure}
    \hfill
    \begin{subfigure}{0.3\textwidth}
        \centering
        \includegraphics[width=\textwidth]{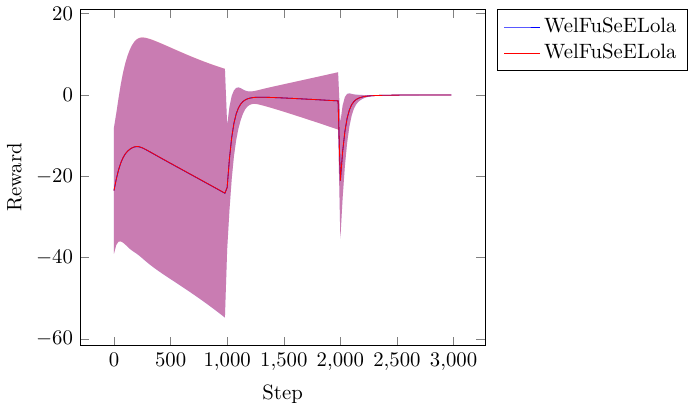}
        \caption{}
        \label{fig:welfuseelola sp chicken game rewards}
    \end{subfigure}
    \hfill
    \begin{subfigure}{0.3\textwidth}
        \centering
        \includegraphics[width=\textwidth]{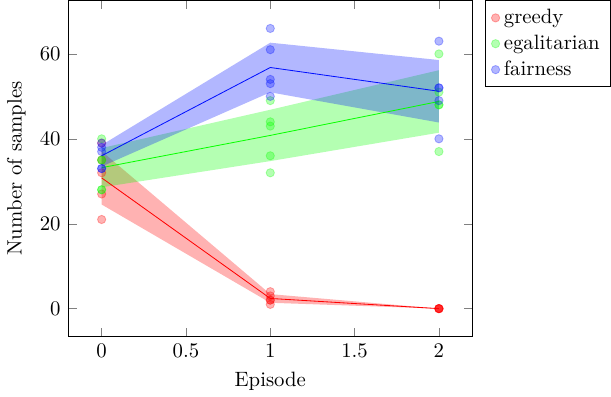}
        \caption{}
        \label{fig:welfuseelola sp chicken game welfare function history}
    \end{subfigure}
    \caption{Comparing ELOLA ($\eta=0.1, \alpha=25$, averaged over 100 trials) with WelFuSeElola (same ELOLA hyperparameters, $e=3, s=1000, b=100$, averaged over five random seeds) against different opponents in Chicken Game. Top row: against NL. Bottom row: self-play. Left column: rewards for ELOLA. Centre column: rewards for WelFuSeElola. Right column: welfare functions chosen by WelFuSeElola in each episode.}
    \label{fig:chicken game sp and nl}
\end{figure*}

\subsection{Welfare Function Search} \label{section:we/welfuse}
The previous subsection demonstrated that the best welfare function to choose in a WE strategy depends on the game and the nature of the opponent, but did not explain how to choose the best welfare function using a practical algorithm, which we introduce in this subsection. The welfare function is effectively an optimisation objective function (given a suitable approximation of the opponent BR function), but rather than searching over the entire continuous space of all possible welfare functions \cite{lu2022discovered}, we instead take a simpler approach and assume access to a finite set of pre-defined welfare functions, such as the three that were considered in Section \ref{section:we/chicken catastrophe}, which have been shown to find mutually desirable NE in a variety of canonical non-coincidental games. We also assume access to an inner OS algorithm (such as ELOLA), which is able to optimise a given choice of welfare function, rather than the usual self-reward. In the OS update step, the opponent is always assumed to maximise self-reward, consistent with the definition of WE in Equations \ref{eq:we x} and \ref{eq:we y}. This leads to the algorithm ``Welfare Function Search'' (WelFuSe), a practical algorithm for adaptively choosing a welfare function from experience, which preserves performance against NL while avoiding catastrophe in self-play. The principle behind WelFuSe is to treat the problem of choosing a welfare function as a discrete bandit problem, which is solved over multiple episodes using a batched variant of posterior sampling \cite{sutton2018reinforcement}, in order to maximise final \emph{self-reward} from each episode. After each episode, all agents' strategies are reset, and a new batch of welfare functions are sampled. The full WelFuSe algorithm is described in Algorithm \ref{alg:welfuse} in the Appendix.

The results from using WelFuSe with ELOLA as the inner OS algorithm (which we jointly refer to as ``WelFuSeElola'') in the Chicken Game are shown in Figure \ref{fig:chicken game sp and nl} in the Appendix. These results demonstrate that against NL, WelFuSeElola quickly learns to reject egalitarian and fairness welfare functions in order to optimise a greedy welfare function (Figure \ref{fig:welfuseelola nl chicken game welfare function history}). This leads to the optimal reward against NL (Figure \ref{fig:welfuseelola nl chicken game rewards}), which is equal to the performance of ELOLA (Figure \ref{fig:elola nl chicken game}). However in self-play (when both agents are WelFuSeElola agents, playing against each other with separate welfare function distributions, while each assumes their opponent takes an approximate NL update step), WelFuSeElola learns to reject the greedy welfare function and instead optimise a mixture of egalitarian and fairness welfare functions (Figure \ref{fig:welfuseelola sp chicken game welfare function history}), leading to maximally fair and egalitarian outcomes for both WelFuSeElola self-play agents (Figure \ref{fig:welfuseelola sp chicken game rewards}). This is much more desirable for both agents than the catastrophe experienced by ELOLA agents in self-play (Figure \ref{fig:elola sp chicken game}), and we therefore conclude that WelFuSeElola has solved the chicken catastrophe.

We end this subsection by emphasising that WelFuSe can be applied given any reasonable choice of inner OS algorithm, and therefore WelFuSe should be considered as an \emph{extension} to existing OS algorithms, rather than a mutually exclusive alternative. As we have shown, WelFuSe learns to select a non-greedy welfare function when it is in the agent's interest to do so. In other cases, WelFuSe learns to reject all other welfare functions and learns to only optimise a greedy welfare function, in which case WelFuSe simply reduces to its inner OS algorithm.

\section{Conclusions}
In Section \ref{section:os stackelberg/solution concept} we showed that Stackelberg strategy profiles provide a sensible solution concept in a variety of games, which we refer to as coincidental games, as well as in the Impossible Market, which contains no NE. In Section \ref{section:os stackelberg/approximation} we showed that Stackelberg strategies represent a unifying framework from which many existing OS algorithms can be derived as approximations. In Section \ref{section:os stackelberg/new os algorithms} we demonstrated the value of this framework by using it to derive new algorithms which have qualitative advantages over previous approaches. For example, SaGa was able to consistently converge in the Impossible Market (unlike many learning algorithms), and SaSa was able to consistently converge to the global optimum in Stag Hunt while avoiding the non-global local optimum. SaSa was also able to strongly dominate NL in IPD in a single episode.

In Section \ref{section:we/nc games} we defined non-coincidental games as games in which the Stackelberg strategy profile is not a NE, which notably includes several canonical matrix games, and illustrates why OS algorithms can fail in self-play in such games. Avoiding catastrophe in self-play is important for any learning algorithm to be safe to deploy in multi-agent systems in the real world. To this end, in Section \ref{section:we/chicken catastrophe} we introduced WE as a generalisation of the Stackelberg framework, which can recover desirable NE solutions in non-coincidental games. We showed that against another WE agent in the Tandem game, choosing a greedy welfare function is a strictly dominated strategy, and therefore in this example both players have a greedy incentive to maximise a non-greedy welfare function. In Section \ref{section:we/welfuse} we introduced WelFuSe as a practical approach to choosing between welfare functions, which was able to preserve the performance of ELOLA against NL, while also avoiding catastrophe in self-play.

Despite these contributions, this work offers alluring directions for future work, such as developing new OS algorithms using more sophisticated approximations to Stackelberg strategies, and more sophisticated approaches for choosing effective welfare functions. Overall, we hope that this work takes us closer towards understanding how to design multi-agent learning algorithms that are both safe and effective in the real world.

\section*{Acknowledgements}
This work was supported by the EPSRC Centre for Doctoral Training in Autonomous Intelligent Machines and Systems (grant number EP/S024050/1). CSDW received generous support by the Cooperative AI Foundation. This work was supported by Armasuisse Science+Technology.

\section*{Impact Statement}
This paper presents contributions which are intended to improve understanding and learning outcomes in general-sum multi-agent systems. We hope that this will contribute to improved outcomes in multi-agent systems in the real world, including for models which learn while interacting with humans. Such models have caused negative real-world outcomes in the past \cite{wolf2017we} due to naive design, and we hope that our contributions may help to design better models which avoid such outcomes in the future.

\bibliographystyle{icml2024}
\bibliography{references}

\appendix

\clearpage
\section{Appendix}
\label{appendix:additional results}

\subsection{Arrogance Penalties And Invariant Welfare Functions} \label{section:we/arrogance penalty}

The egalitarian and fairness welfare functions provide mutually desirable NE in a variety of non-coincidental games, however they share a common disadvantage, which is that unlike BR functions and Stackelberg strategies, they are not invariant to shifting and positive scaling of either player's reward function. We can address this issue by introducing ``arrogance penalties''. When a player decides to play the Stackelberg strategy (drive straight) in the chicken game, they effectively assume that the opponent will play a BR (chicken out) to the Stackelberg strategy, in which case the former player expects a reward of $+1$. We refer to the reward for either player when they play a Stackelberg strategy and the opponent plays a BR as the ``Stackelberg baseline''. When both players choose Stackelberg strategies in the chicken game and mutually drive straight, they both receive a reward of $-100$, which is $101$ less than the Stackelberg baseline. This discrepency between expected and received rewards for each player is effectively a penalty for arrogantly assuming that they could ``call the shots'' while the opponent would ``fall in line''. We therefore refer to the difference between the Stackelberg baseline and the reward from the Stackelberg strategy profile as the ``arrogance penalty'' for each player respectively. Following this intuition, we can define the shift-normalised reward functions $\pi^x(x, y)$ and $\pi^y(x, y)$ for any strategy profile as the reward that each player receives relative to their Stackelberg baseline:

\begin{align}
    \pi^x(x, y) &= R^x(x, y) - R^x \Bigl(  x^*, y^*\left( x^* \right)  \Bigr) \label{eq:shift normalised reward x} \\
    \pi^y(x, y) &= R^y(x, y) - R^y \Bigl(  x^*\left( y^* \right), y^*  \Bigr) \label{eq:shift normalised reward y}
\end{align}

We note that for either player, if their opponent will always play a BR, then the shift-normalised reward function for the former player is negative everywhere except when they play a Stackelberg strategy, at which point the shift-normalised reward function is zero (this follows from the definition of the Stackelberg strategies in equations \ref{eq:stackelberg x} and \ref{eq:stackelberg y}). We also note that, because both players' BR functions and Stackelberg strategies are invariant to shifting of either player's reward function, so too are $\pi^x(x, y)$ and $\pi^y(x, y)$, because any shift in $R^x(x, y)$ or $R^y(x, y)$ would be cancelled out between the two terms in equations \ref{eq:shift normalised reward x} and \ref{eq:shift normalised reward y} respectively. This allows us to define the shift-normalised egalitarian welfare function as follows:

\begin{equation}
    \pi_{\textrm{egalitarian}}(x, y) = \min\Bigl( \pi^x(x, y), \pi^y(x, y) \Bigr) \label{eq:shift normalised egalitarian} \\
\end{equation}

We note that in the special case of coincidental games, when both players use the shift-normalised egalitarian welfare function, the resulting WE strategy profile is always the Stackelberg strategy profile. The shift-normalised egalitarian welfare function is invariant to shifts in either player's reward function, but it is not invariant to scaling. To address this issue, for non-coincidental games, we can introduce the affinely-normalised reward functions (equal to the shift-normalised reward functions normalised by the absolute arrogance penalties) and affinely-normalised egalitarian welfare function as follows:

\begin{align}
    \bar{\pi}^x(x, y) &= \frac{R^x(x, y) - R^x \Bigl(  x^*, y^*\left( x^* \right)  \Bigr)}{\left\vert R^x \Bigl(  x^*, y^*\left( x^* \right)  \Bigr) - R^x \Bigl(  x^*, y^*  \Bigr)\right\vert} \\
    \bar{\pi}^y(x, y) &= \frac{R^y(x, y) - R^y \Bigl(  x^*\left( y^* \right), y^*  \Bigr)}{\left\vert R^y \Bigl(  x^*\left( y^* \right), y^*  \Bigr) - R^y \Bigl(  x^*, y^*  \Bigr)\right\vert} \\
    \bar{\pi}_{\textrm{egalitarian}}(x, y) &= \min\Bigl( \bar{\pi}^x(x, y), \bar{\pi}^y(x, y) \Bigr) \label{eq:affinely normalised egalitarian}
\end{align}

The affinely-normalised fairness welfare function can be defined analogously. The affinely-normalised egalitarian welfare function (Equation \ref{eq:affinely normalised egalitarian}) is invariant to shifting and positive scaling of either player's reward function. Therefore, if it is used by both players as the welfare function in a WE strategy profile, then the strategies chosen by both players are also invariant to shifting and positive scaling of either player's reward function. Furthermore, in symmetric games, the Stackelberg baseline and arrogance penalty are necessarily symmetric for both players, which implies that the affinely-normalised egalitarian WE strategy profile is simply equivalent to the egalitarian WE strategy profile in symmetric games.

What is offered by the affinely-normalised egalitarian welfare function in terms of mathematical appeal is compensated by its greater computational burden in asymmetric games, because it depends on knowing the Stackelberg strategies for both players and their respective BRs before it can be used to calculate a WE strategy. This would limit the applicability to online learning settings, although it could be used if there was the opportunity to approximate Stackelberg strategies and BRs offline before the start of online learning. This approach would also be suitable to offline learning settings, in which the Stackelberg strategies, BRs, and WE strategies could be learned sequentially. In general, we leave the development of invariant welfare functions which are more conducive to efficient practical online implementation as an interesting direction for future work.

\begin{table*}[h!]
    \centering
    \begin{subtable}[t]{0.45\textwidth}
        \centering
        \caption{Payoff table for Prisoners' Dilemma}
        \begin{tabular}[t]{l|c c}
            \hline
            & \textcolor{violet}{Cooperate}                         & \textcolor{violet}{Defect}                              \\
            \hline
            \textcolor{blue}{Cooperate}                             & (\textcolor{blue}{-1.0}, \textcolor{violet}{-1.0})      & (\textcolor{blue}{-3.0}, \textcolor{violet}{0.0})       \\
            \textcolor{blue}{Defect}                                & (\textcolor{blue}{0.0}, \textcolor{violet}{-3.0})       & (\textcolor{blue}{-2.0}, \textcolor{violet}{-2.0})      \\
            \hline
        \end{tabular}
        \label{table:PrisonersDilemma payoffs}
    \end{subtable}
    \hfill
    \begin{subtable}[t]{0.45\textwidth}
        \centering
        \caption{Payoff table for Matching Pennies}
        \begin{tabular}[t]{l|c c}
            \hline
            & \textcolor{violet}{Heads}                             & \textcolor{violet}{Tails}                               \\
            \hline
            \textcolor{blue}{Heads}                                 & (\textcolor{blue}{1.0}, \textcolor{violet}{-1.0})       & (\textcolor{blue}{-1.0}, \textcolor{violet}{1.0})       \\
            \textcolor{blue}{Tails}                                 & (\textcolor{blue}{-1.0}, \textcolor{violet}{1.0})       & (\textcolor{blue}{1.0}, \textcolor{violet}{-1.0})       \\
            \hline
        \end{tabular}
        \label{table:MatchingPennies payoffs}
    \end{subtable}
    \newline
    \newline
    \newline
    \newline
    \begin{subtable}[t]{0.4\textwidth}
        \centering
        \caption{Payoff table for Stag Hunt}
        \begin{tabular}[t]{l|c c}
            \hline
            & \textcolor{violet}{Stag}                              & \textcolor{violet}{Hare}                                \\
            \hline
            \textcolor{blue}{Stag}                                  & (\textcolor{blue}{10.0}, \textcolor{violet}{10.0})      & (\textcolor{blue}{1.0}, \textcolor{violet}{8.0})        \\
            \textcolor{blue}{Hare}                                  & (\textcolor{blue}{8.0}, \textcolor{violet}{1.0})        & (\textcolor{blue}{5.0}, \textcolor{violet}{5.0})        \\
            \hline
        \end{tabular}
        \label{table:StagHunt payoffs}
    \end{subtable}
    \hfill
    \begin{subtable}[t]{0.55\textwidth}
        \centering
        \caption{Payoff table for Chicken Game}
        \begin{tabular}[t]{l|c c}
            \hline
            & \textcolor{violet}{Chicken out}                       & \textcolor{violet}{Drive straight}                      \\
            \hline
            \textcolor{blue}{Chicken out}                           & (\textcolor{blue}{0.0}, \textcolor{violet}{0.0})        & (\textcolor{blue}{-1.0}, \textcolor{violet}{1.0})       \\
            \textcolor{blue}{Drive straight}                        & (\textcolor{blue}{1.0}, \textcolor{violet}{-1.0})       & (\textcolor{blue}{-100.0}, \textcolor{violet}{-100.0})  \\
            \hline
        \end{tabular}
        \label{table:ChickenGame payoffs}
    \end{subtable}
    \newline
    \newline
    \newline
    \newline
    \begin{subtable}[t]{0.4\textwidth}
        \centering
        \caption{Payoff table for Bach Or Stravinsky (BOS)}
        \begin{tabular}[t]{l|c c}
            \hline
            & \textcolor{violet}{Bach}                              & \textcolor{violet}{Stravinsky}                          \\
            \hline
            \textcolor{blue}{Bach}                                  & (\textcolor{blue}{2.0}, \textcolor{violet}{1.0})        & (\textcolor{blue}{0.0}, \textcolor{violet}{0.0})        \\
            \textcolor{blue}{Stravinsky}                            & (\textcolor{blue}{0.0}, \textcolor{violet}{0.0})        & (\textcolor{blue}{1.0}, \textcolor{violet}{2.0})        \\
            \hline
        \end{tabular}
        \label{table:CoordinationGame payoffs}
    \end{subtable}
    \hfill
    \begin{subtable}[t]{0.55\textwidth}
        \centering
        \caption{Payoff table for Baby Chicken Game}
        \begin{tabular}[t]{l|c c}
            \hline
            & \textcolor{violet}{Chicken out}                       & \textcolor{violet}{Drive straight}                      \\
            \hline
            \textcolor{blue}{Chicken out}                           & (\textcolor{blue}{0.0}, \textcolor{violet}{0.0})        & (\textcolor{blue}{-1.0}, \textcolor{violet}{1.0})       \\
            \textcolor{blue}{Drive straight}                        & (\textcolor{blue}{1.0}, \textcolor{violet}{-1.0})       & (\textcolor{blue}{-3.0}, \textcolor{violet}{-3.0})      \\
            \hline
        \end{tabular}
        \label{table:BabyChickenGame payoffs}
    \end{subtable}
    \newline
    \newline
    \newline
    \newline
    \begin{subtable}[t]{0.45\textwidth}
        \centering
        \caption{Payoff table for Awkward Game}
        \begin{tabular}[t]{l|c c}
            \hline
            & \textcolor{violet}{Cooperate}                         & \textcolor{violet}{Defect}                              \\
            \hline
            \textcolor{blue}{Cooperate}                             & (\textcolor{blue}{3.0}, \textcolor{violet}{1.0})        & (\textcolor{blue}{1.0}, \textcolor{violet}{3.0})        \\
            \textcolor{blue}{Defect}                                & (\textcolor{blue}{2.0}, \textcolor{violet}{5.0})        & (\textcolor{blue}{4.0}, \textcolor{violet}{2.0})        \\
            \hline
        \end{tabular}
        \label{table:AwkwardGame payoffs}
    \end{subtable}
    \hfill
    \begin{subtable}[t]{0.45\textwidth}
        \centering
        \caption{Payoff table for Eagle Game}
        \begin{tabular}[t]{l|c c}
            \hline
            & \textcolor{violet}{Cooperate}                         & \textcolor{violet}{Defect}                              \\
            \hline
            \textcolor{blue}{Cooperate}                             & (\textcolor{blue}{4.0}, \textcolor{violet}{1.0})        & (\textcolor{blue}{-4.0}, \textcolor{violet}{-1.0})      \\
            \textcolor{blue}{Defect}                                & (\textcolor{blue}{-2.0}, \textcolor{violet}{-3.0})      & (\textcolor{blue}{2.0}, \textcolor{violet}{3.0})        \\
            \hline
        \end{tabular}
        \label{table:EagleGame payoffs}
    \end{subtable}
    \caption{Payoff tables for matrix games}
    \label{table:matrix game payoff tables}
\end{table*}

\begin{algorithm*}
    \caption{Welfare Function Search (WelFuSe)}\label{alg:welfuse}
        \begin{algorithmic}[1]
            \REQUIRE Set of welfare functions $\mathcal{W} = \left\{w_1, \dots, w_{\vert \mathcal{W} \vert}\right\}$, number of episodes $e$, steps per episode $s$, batch size $b$, inner OS algorithm $\texttt{os}(w_k, s)$
            \FOR{$j \in (1, \dots, b)$}
                \STATE $k \sim \mathcal{U}(\{1, \dots, \vert \mathcal{W} \vert\})$ \textcolor{gray}{\;\;\; // Sample initial welfare functions uniformly}
                \STATE $w^{(1, j)} \gets w_k$
            \ENDFOR
            \FOR{$i \in (1, \dots, e)$}
                \FOR{$j \in (1, \dots, b)$}
                    \STATE $\texttt{os}(w^{(i, j)}, s)$ \textcolor{gray}{\;\;\; // Optimise welfare function using OS}
                    \STATE Observe final reward $r^{(i, j)}$
                \ENDFOR
                \STATE Reset all agents' parameters
                \STATE \textcolor{gray}{// Consider each batch-index}
                \FOR{$j \in (1, \dots, b)$}
                    \STATE \textcolor{gray}{// Consider each welfare function $w_k$}
                    \FOR{$k \in (1, \dots, \vert \mathcal{W} \vert)$}
                        \STATE $m_k \sim \mathcal{U}(\{ n \in \{1, \dots, b\} : w^{(i, n)} = w_k \})$ \textcolor{gray}{\;\;\; // Sample batch-index which used $w_k$}
                    \ENDFOR
                    \STATE $\hat{k} \gets \underset{k}{\mathrm{argmax}}\left[ r^{(i, m_k)} \right]$ \textcolor{gray}{\;\;\; // Select welfare function with best sampled reward}
                    \STATE $w^{(i+1, j)} \gets w_{\hat{k}}$ \textcolor{gray}{\;\;\; // Set welfare function for next episode}
                \ENDFOR
            \ENDFOR
        \end{algorithmic}
    \end{algorithm*}

\clearpage
\begin{figure*}
    \centering
    \captionsetup{justification=centering}
    \begin{subfigure}[t]{0.45\textwidth}
        \centering
        \includegraphics[width=\textwidth]{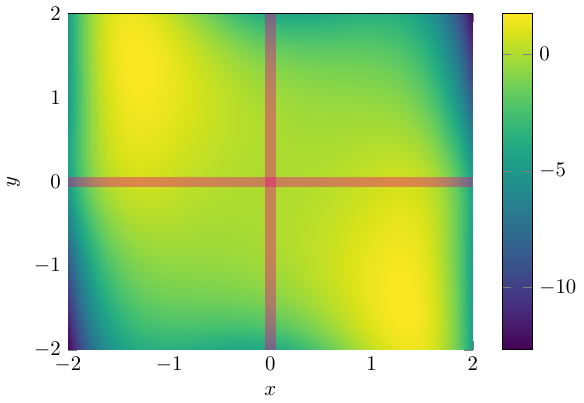}
        \caption{Reward surface for $x$}
        \label{fig:fig:ImpossibleMarket_greedy reward_surface_x}
    \end{subfigure}
    \hfill
    \begin{subfigure}[t]{0.45\textwidth}
        \centering
        \includegraphics[width=\textwidth]{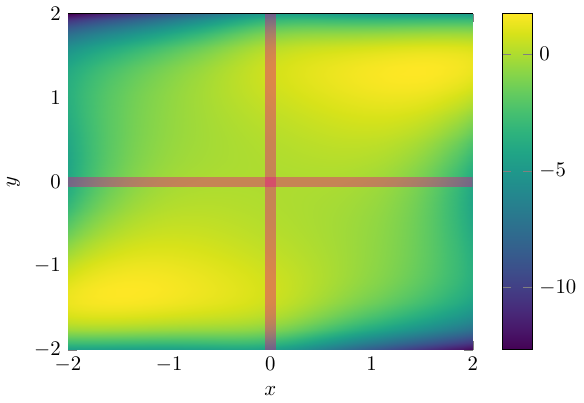}
        \caption{Reward surface for $y$}
        \label{fig:fig:ImpossibleMarket_greedy reward_surface_y}
    \end{subfigure}
    \newline
    \begin{subfigure}[t]{0.45\textwidth}
        \centering
        \includegraphics[width=\textwidth]{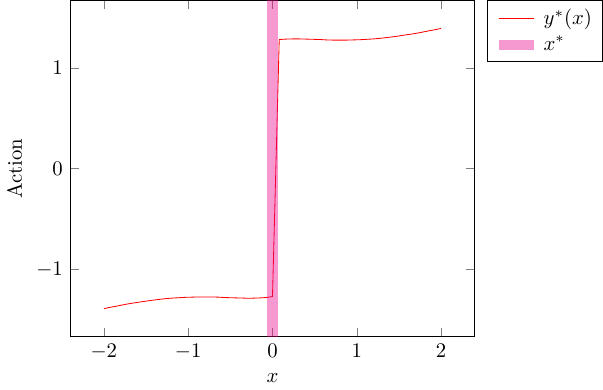}
        \caption{BR for $y$ as a function of $x$}
        \label{fig:fig:ImpossibleMarket_greedy y_br}
    \end{subfigure}
    \hfill
    \begin{subfigure}[t]{0.45\textwidth}
        \centering
        \includegraphics[width=\textwidth]{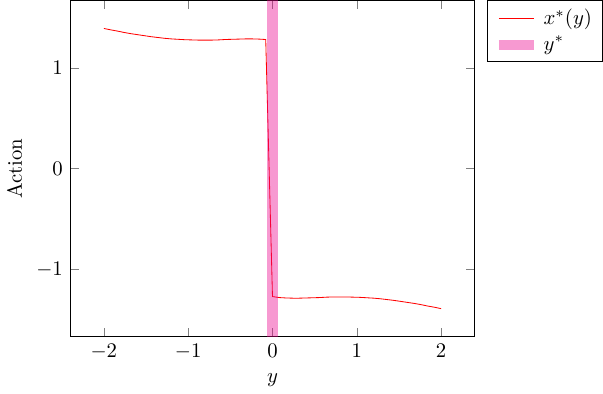}
        \caption{BR for $x$ as a function of $y$}
        \label{fig:fig:ImpossibleMarket_greedy x_br}
    \end{subfigure}
    \newline
    \begin{subfigure}[t]{0.45\textwidth}
        \centering
        \includegraphics[width=\textwidth]{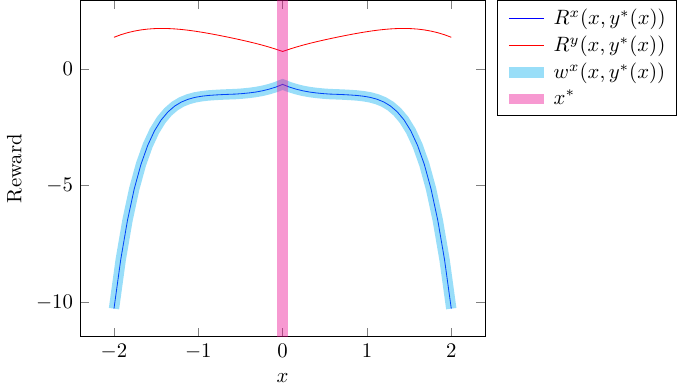}
        \caption{Rewards as a function of $x$}
        \label{fig:fig:ImpossibleMarket_greedy br_rewards_x}
    \end{subfigure}
    \hfill
    \begin{subfigure}[t]{0.45\textwidth}
        \centering
        \includegraphics[width=\textwidth]{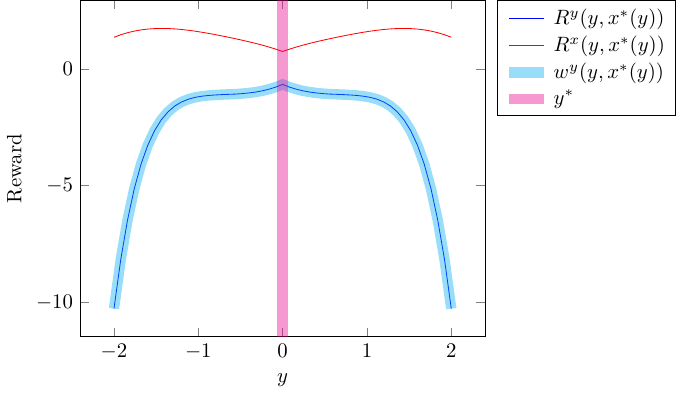}
        \caption{Rewards as a function of $y$}
        \label{fig:fig:ImpossibleMarket_greedy br_rewards_y}
    \end{subfigure}
    \caption{Stackelberg strategy profile (Greedy WE) for ImpossibleMarket \\ $x^* = 0.000, y^* = 0.000, R^x = -0.000, R^y = -0.000$}
    \label{fig:ImpossibleMarket_greedy}
\end{figure*}

\begin{figure*}
    \centering
    \captionsetup{justification=centering}
    \begin{subfigure}[t]{0.45\textwidth}
        \centering
        \includegraphics[width=\textwidth]{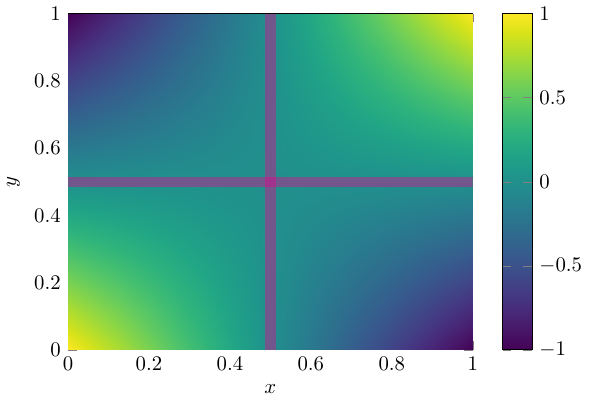}
        \caption{Reward surface for $x$}
        \label{fig:fig:MatchingPennies_greedy reward_surface_x}
    \end{subfigure}
    \hfill
    \begin{subfigure}[t]{0.45\textwidth}
        \centering
        \includegraphics[width=\textwidth]{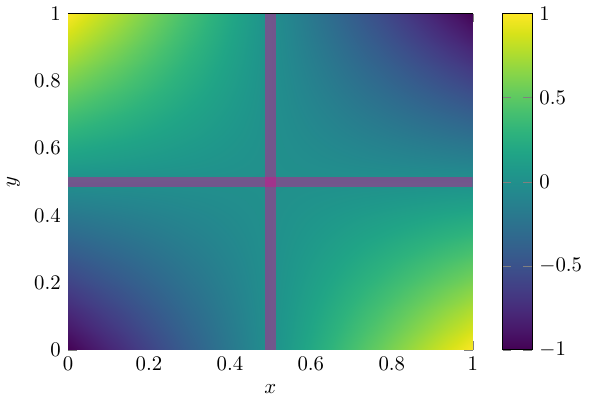}
        \caption{Reward surface for $y$}
        \label{fig:fig:MatchingPennies_greedy reward_surface_y}
    \end{subfigure}
    \newline
    \begin{subfigure}[t]{0.45\textwidth}
        \centering
        \includegraphics[width=\textwidth]{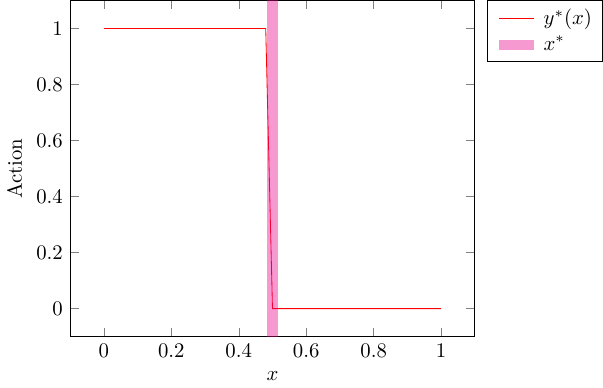}
        \caption{BR for $y$ as a function of $x$}
        \label{fig:fig:MatchingPennies_greedy y_br}
    \end{subfigure}
    \hfill
    \begin{subfigure}[t]{0.45\textwidth}
        \centering
        \includegraphics[width=\textwidth]{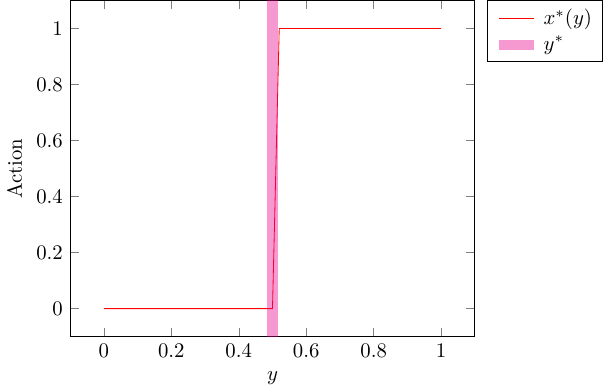}
        \caption{BR for $x$ as a function of $y$}
        \label{fig:fig:MatchingPennies_greedy x_br}
    \end{subfigure}
    \newline
    \begin{subfigure}[t]{0.45\textwidth}
        \centering
        \includegraphics[width=\textwidth]{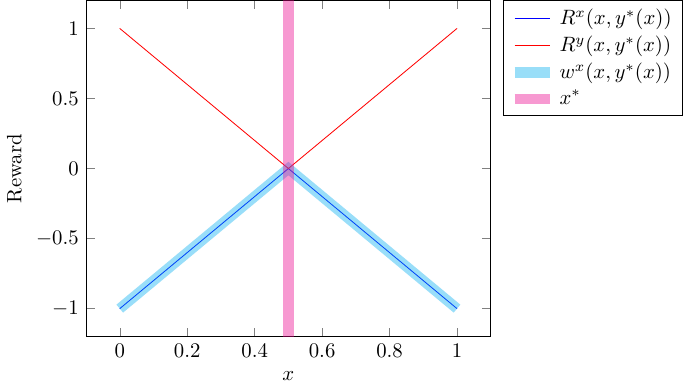}
        \caption{Rewards as a function of $x$}
        \label{fig:fig:MatchingPennies_greedy br_rewards_x}
    \end{subfigure}
    \hfill
    \begin{subfigure}[t]{0.45\textwidth}
        \centering
        \includegraphics[width=\textwidth]{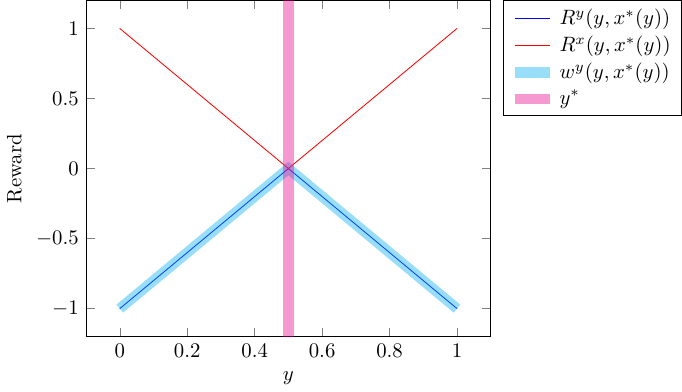}
        \caption{Rewards as a function of $y$}
        \label{fig:fig:MatchingPennies_greedy br_rewards_y}
    \end{subfigure}
    \caption{Stackelberg strategy profile (Greedy WE) for MatchingPennies \\ $x^* = 0.500, y^* = 0.500, R^x = 0.000, R^y = 0.000$}
    \label{fig:MatchingPennies_greedy}
\end{figure*}

\begin{figure*}
    \centering
    \captionsetup{justification=centering}
    \begin{subfigure}[t]{0.45\textwidth}
        \centering
        \includegraphics[width=\textwidth]{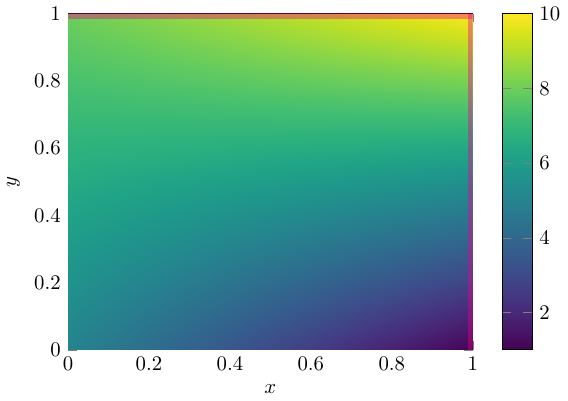}
        \caption{Reward surface for $x$}
        \label{fig:fig:StagHunt_greedy reward_surface_x}
    \end{subfigure}
    \hfill
    \begin{subfigure}[t]{0.45\textwidth}
        \centering
        \includegraphics[width=\textwidth]{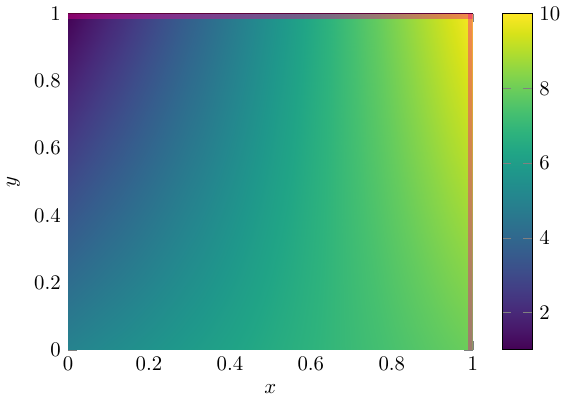}
        \caption{Reward surface for $y$}
        \label{fig:fig:StagHunt_greedy reward_surface_y}
    \end{subfigure}
    \newline
    \begin{subfigure}[t]{0.45\textwidth}
        \centering
        \includegraphics[width=\textwidth]{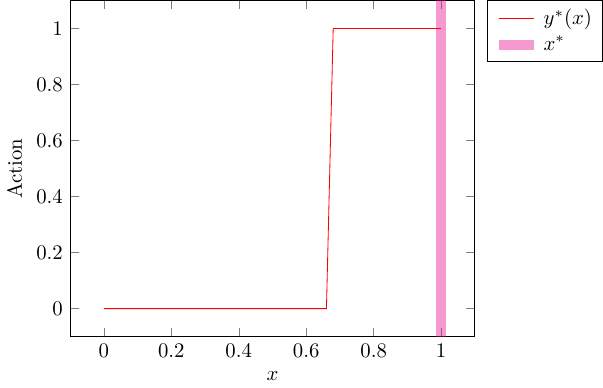}
        \caption{BR for $y$ as a function of $x$}
        \label{fig:fig:StagHunt_greedy y_br}
    \end{subfigure}
    \hfill
    \begin{subfigure}[t]{0.45\textwidth}
        \centering
        \includegraphics[width=\textwidth]{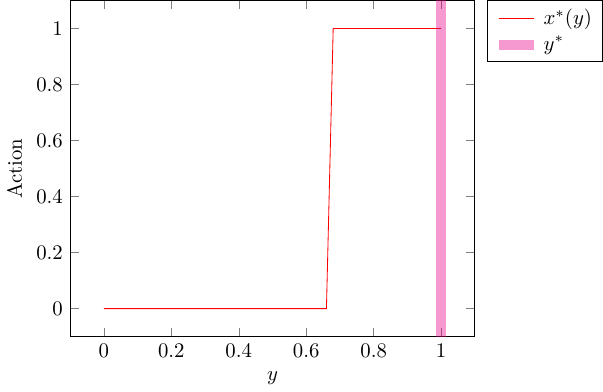}
        \caption{BR for $x$ as a function of $y$}
        \label{fig:fig:StagHunt_greedy x_br}
    \end{subfigure}
    \newline
    \begin{subfigure}[t]{0.45\textwidth}
        \centering
        \includegraphics[width=\textwidth]{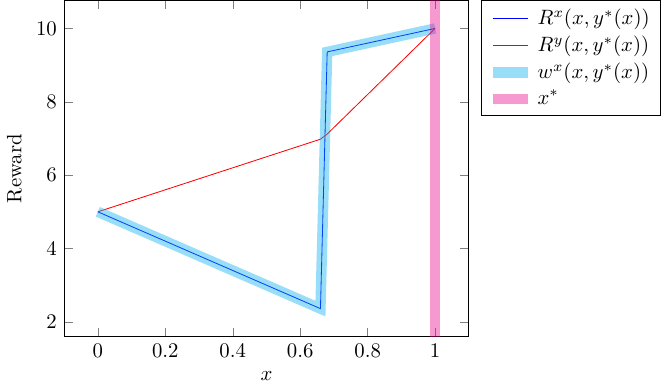}
        \caption{Rewards as a function of $x$}
        \label{fig:fig:StagHunt_greedy br_rewards_x}
    \end{subfigure}
    \hfill
    \begin{subfigure}[t]{0.45\textwidth}
        \centering
        \includegraphics[width=\textwidth]{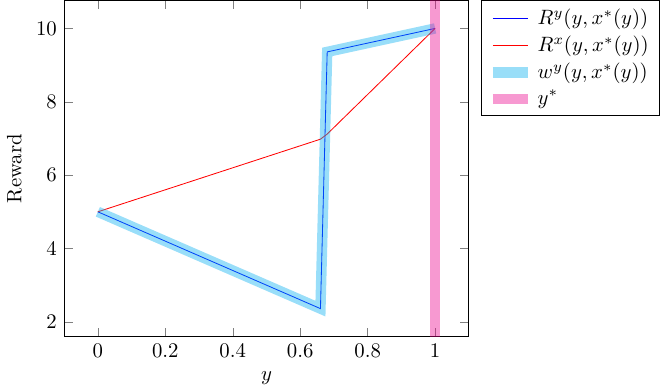}
        \caption{Rewards as a function of $y$}
        \label{fig:fig:StagHunt_greedy br_rewards_y}
    \end{subfigure}
    \caption{Stackelberg strategy profile (Greedy WE) for StagHunt \\ $x^* = 1.000, y^* = 1.000, R^x = 10.000, R^y = 10.000$}
    \label{fig:StagHunt_greedy}
\end{figure*}

\begin{figure*}
    \centering
    \captionsetup{justification=centering}
    \begin{subfigure}[t]{0.45\textwidth}
        \centering
        \includegraphics[width=\textwidth]{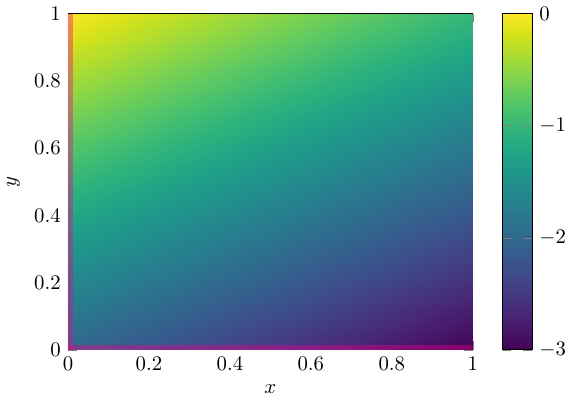}
        \caption{Reward surface for $x$}
        \label{fig:fig:PrisonersDilemma_greedy reward_surface_x}
    \end{subfigure}
    \hfill
    \begin{subfigure}[t]{0.45\textwidth}
        \centering
        \includegraphics[width=\textwidth]{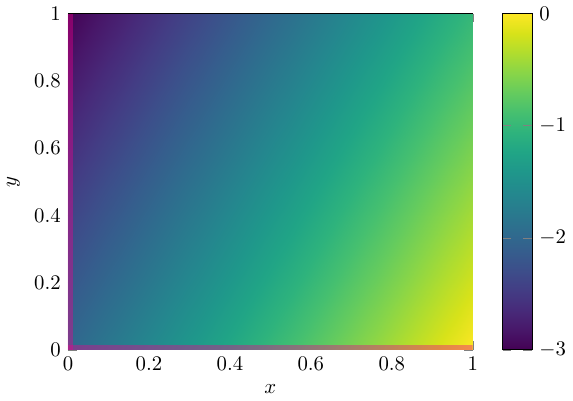}
        \caption{Reward surface for $y$}
        \label{fig:fig:PrisonersDilemma_greedy reward_surface_y}
    \end{subfigure}
    \newline
    \begin{subfigure}[t]{0.45\textwidth}
        \centering
        \includegraphics[width=\textwidth]{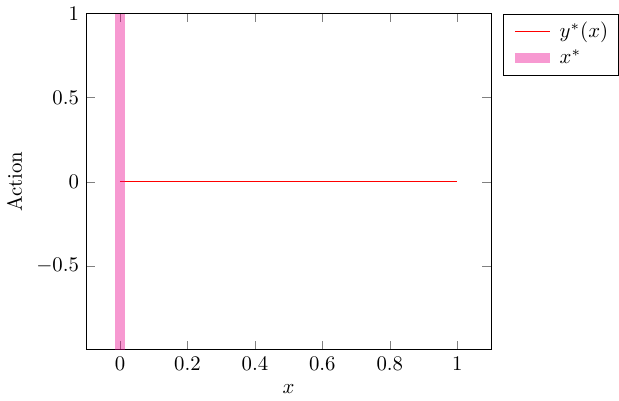}
        \caption{BR for $y$ as a function of $x$}
        \label{fig:fig:PrisonersDilemma_greedy y_br}
    \end{subfigure}
    \hfill
    \begin{subfigure}[t]{0.45\textwidth}
        \centering
        \includegraphics[width=\textwidth]{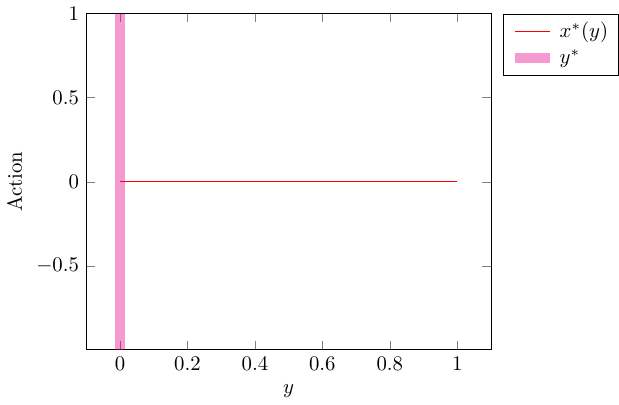}
        \caption{BR for $x$ as a function of $y$}
        \label{fig:fig:PrisonersDilemma_greedy x_br}
    \end{subfigure}
    \newline
    \begin{subfigure}[t]{0.45\textwidth}
        \centering
        \includegraphics[width=\textwidth]{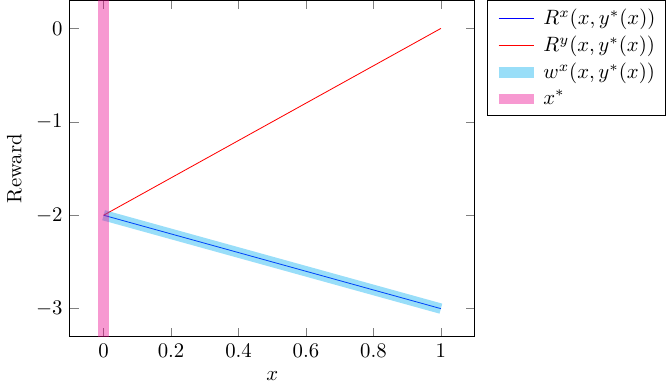}
        \caption{Rewards as a function of $x$}
        \label{fig:fig:PrisonersDilemma_greedy br_rewards_x}
    \end{subfigure}
    \hfill
    \begin{subfigure}[t]{0.45\textwidth}
        \centering
        \includegraphics[width=\textwidth]{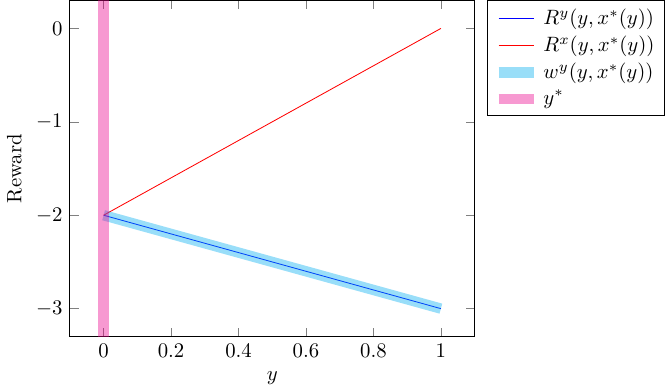}
        \caption{Rewards as a function of $y$}
        \label{fig:fig:PrisonersDilemma_greedy br_rewards_y}
    \end{subfigure}
    \caption{Stackelberg strategy profile (Greedy WE) for PrisonersDilemma \\ $x^* = 0.000, y^* = 0.000, R^x = -2.000, R^y = -2.000$}
    \label{fig:PrisonersDilemma_greedy}
\end{figure*}

\begin{figure*}
    \centering
    \captionsetup{justification=centering}
    \begin{subfigure}[t]{0.45\textwidth}
        \centering
        \includegraphics[width=\textwidth]{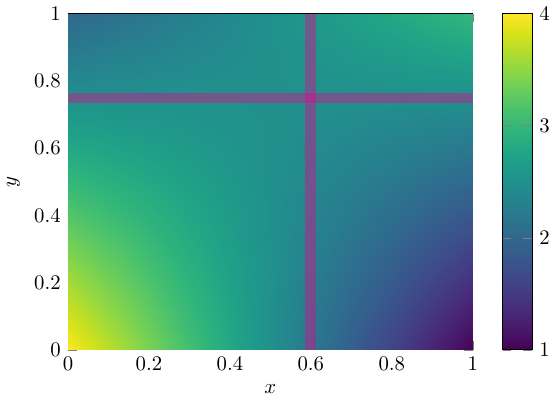}
        \caption{Reward surface for $x$}
        \label{fig:fig:AwkwardGame_greedy reward_surface_x}
    \end{subfigure}
    \hfill
    \begin{subfigure}[t]{0.45\textwidth}
        \centering
        \includegraphics[width=\textwidth]{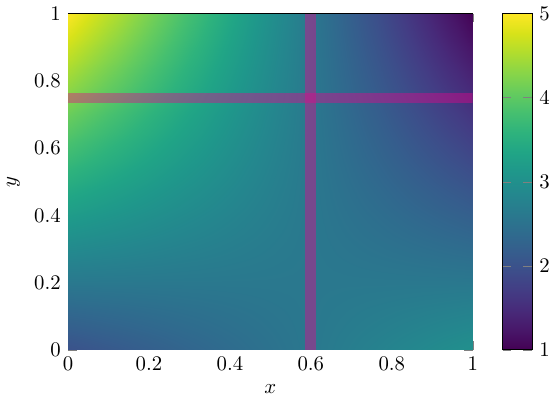}
        \caption{Reward surface for $y$}
        \label{fig:fig:AwkwardGame_greedy reward_surface_y}
    \end{subfigure}
    \newline
    \begin{subfigure}[t]{0.45\textwidth}
        \centering
        \includegraphics[width=\textwidth]{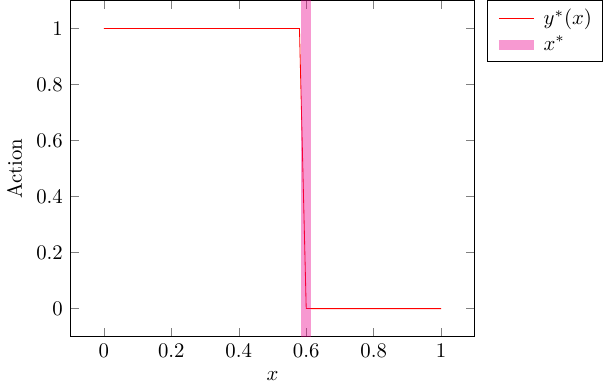}
        \caption{BR for $y$ as a function of $x$}
        \label{fig:fig:AwkwardGame_greedy y_br}
    \end{subfigure}
    \hfill
    \begin{subfigure}[t]{0.45\textwidth}
        \centering
        \includegraphics[width=\textwidth]{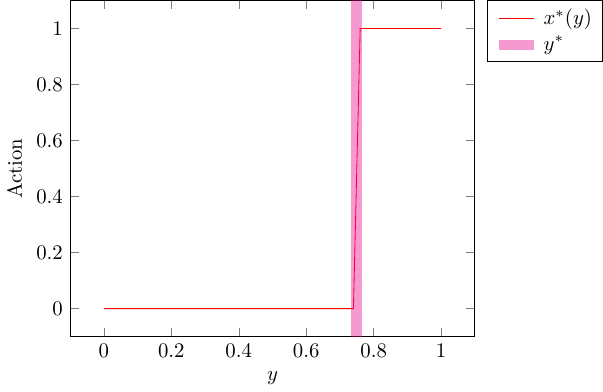}
        \caption{BR for $x$ as a function of $y$}
        \label{fig:fig:AwkwardGame_greedy x_br}
    \end{subfigure}
    \newline
    \begin{subfigure}[t]{0.45\textwidth}
        \centering
        \includegraphics[width=\textwidth]{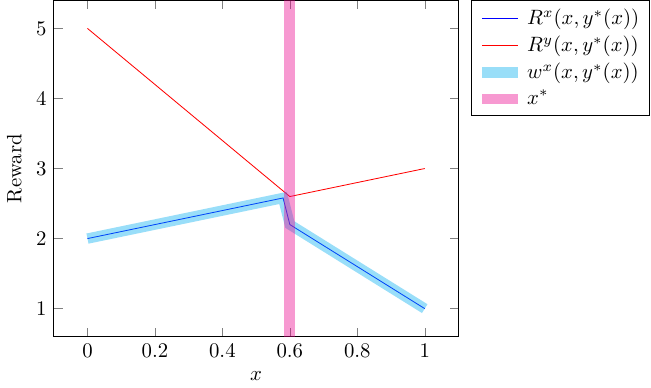}
        \caption{Rewards as a function of $x$}
        \label{fig:fig:AwkwardGame_greedy br_rewards_x}
    \end{subfigure}
    \hfill
    \begin{subfigure}[t]{0.45\textwidth}
        \centering
        \includegraphics[width=\textwidth]{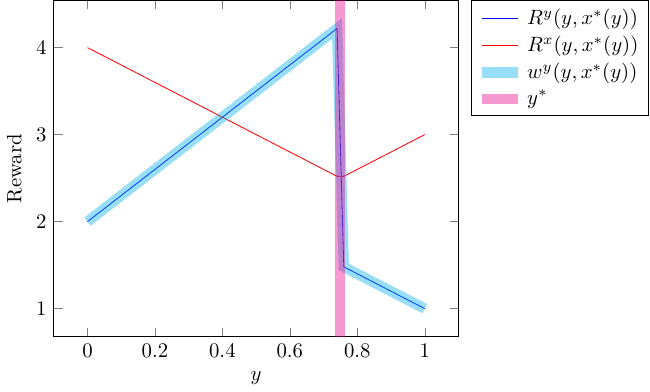}
        \caption{Rewards as a function of $y$}
        \label{fig:fig:AwkwardGame_greedy br_rewards_y}
    \end{subfigure}
    \caption{Stackelberg strategy profile (Greedy WE) for AwkwardGame \\ $x^* = 0.599, y^* = 0.749, R^x = 2.500, R^y = 2.601$}
    \label{fig:AwkwardGame_greedy}
\end{figure*}

\begin{figure*}
    \centering
    \captionsetup{justification=centering}
    \begin{subfigure}[t]{0.45\textwidth}
        \centering
        \includegraphics[width=\textwidth]{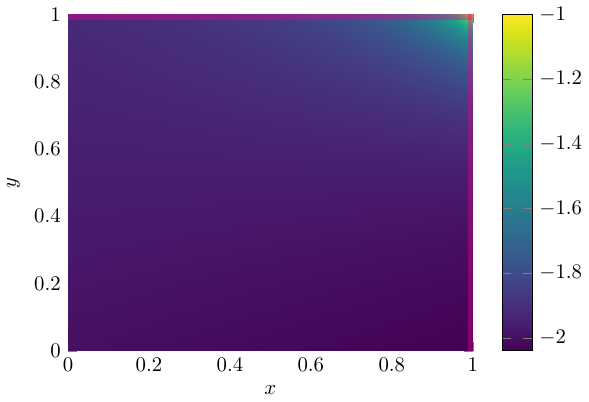}
        \caption{Reward surface for $x$}
        \label{fig:fig:IpdTftAlldMix_greedy reward_surface_x}
    \end{subfigure}
    \hfill
    \begin{subfigure}[t]{0.45\textwidth}
        \centering
        \includegraphics[width=\textwidth]{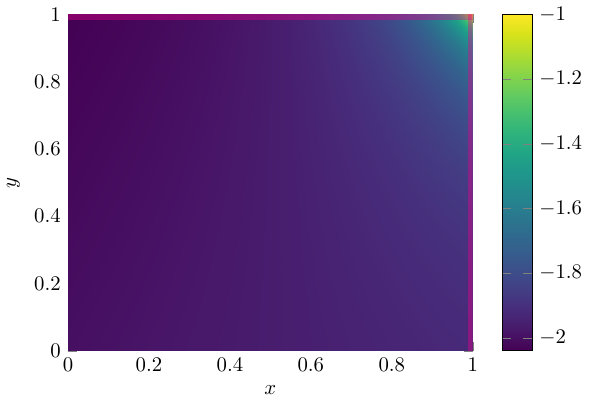}
        \caption{Reward surface for $y$}
        \label{fig:fig:IpdTftAlldMix_greedy reward_surface_y}
    \end{subfigure}
    \newline
    \begin{subfigure}[t]{0.45\textwidth}
        \centering
        \includegraphics[width=\textwidth]{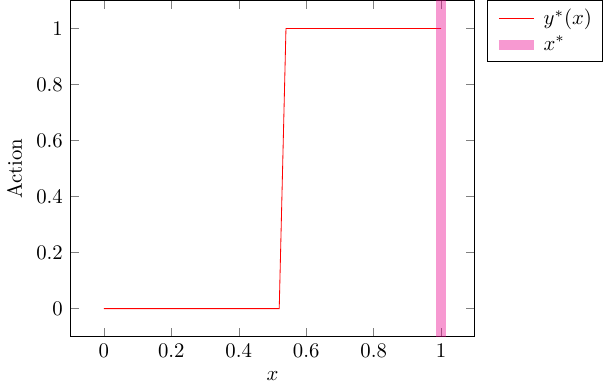}
        \caption{BR for $y$ as a function of $x$}
        \label{fig:fig:IpdTftAlldMix_greedy y_br}
    \end{subfigure}
    \hfill
    \begin{subfigure}[t]{0.45\textwidth}
        \centering
        \includegraphics[width=\textwidth]{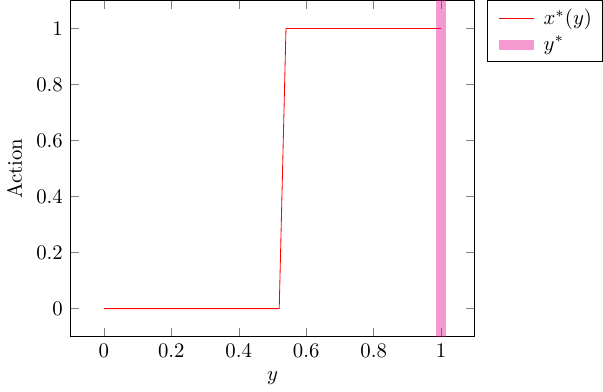}
        \caption{BR for $x$ as a function of $y$}
        \label{fig:fig:IpdTftAlldMix_greedy x_br}
    \end{subfigure}
    \newline
    \begin{subfigure}[t]{0.45\textwidth}
        \centering
        \includegraphics[width=\textwidth]{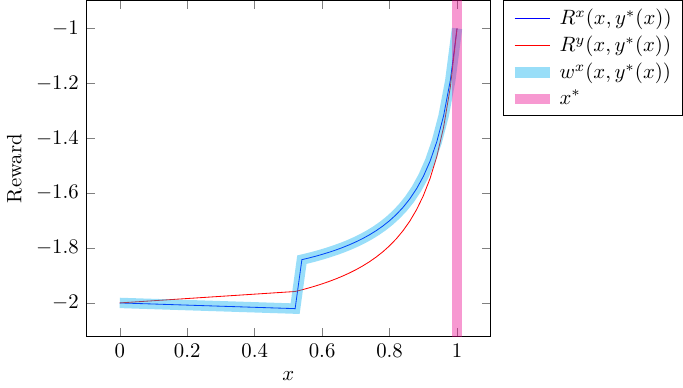}
        \caption{Rewards as a function of $x$}
        \label{fig:fig:IpdTftAlldMix_greedy br_rewards_x}
    \end{subfigure}
    \hfill
    \begin{subfigure}[t]{0.45\textwidth}
        \centering
        \includegraphics[width=\textwidth]{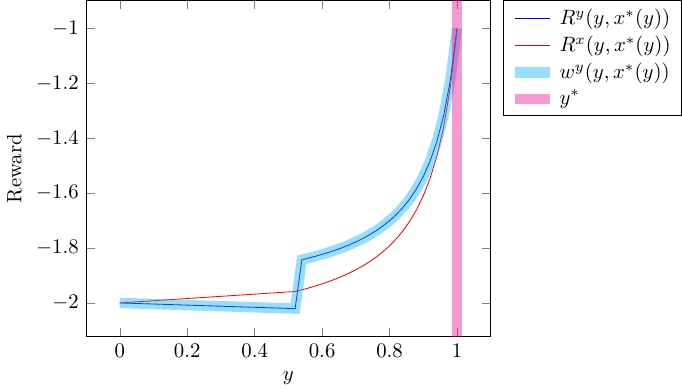}
        \caption{Rewards as a function of $y$}
        \label{fig:fig:IpdTftAlldMix_greedy br_rewards_y}
    \end{subfigure}
    \caption{Stackelberg strategy profile (Greedy WE) for IpdTftAlldMix \\ $x^* = 1.000, y^* = 1.000, R^x = -1.000, R^y = -1.000$}
    \label{fig:IpdTftAlldMix_greedy}
\end{figure*}

\begin{figure*}
    \centering
    \captionsetup{justification=centering}
    \begin{subfigure}[t]{0.45\textwidth}
        \centering
        \includegraphics[width=\textwidth]{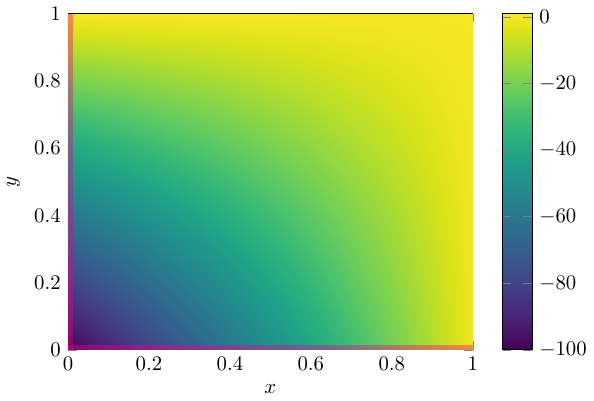}
        \caption{Reward surface for $x$}
        \label{fig:fig:ChickenGame_greedy reward_surface_x}
    \end{subfigure}
    \hfill
    \begin{subfigure}[t]{0.45\textwidth}
        \centering
        \includegraphics[width=\textwidth]{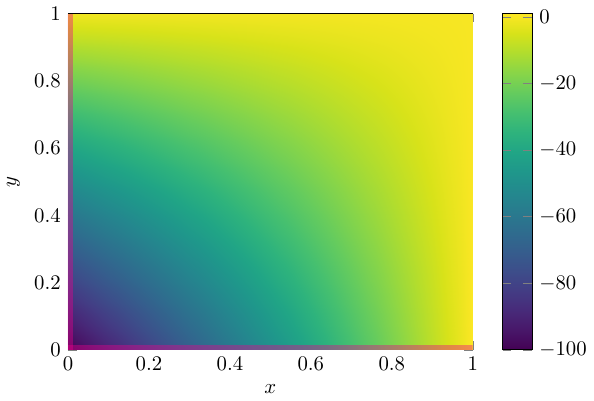}
        \caption{Reward surface for $y$}
        \label{fig:fig:ChickenGame_greedy reward_surface_y}
    \end{subfigure}
    \newline
    \begin{subfigure}[t]{0.45\textwidth}
        \centering
        \includegraphics[width=\textwidth]{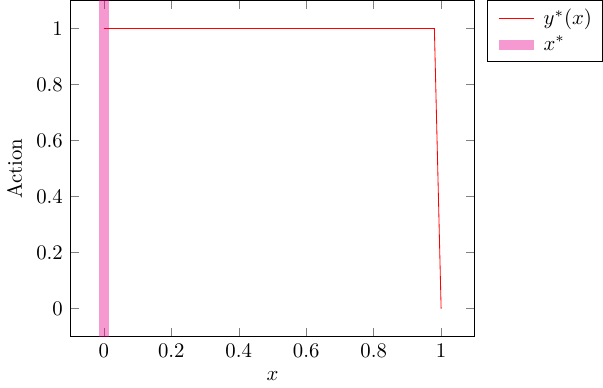}
        \caption{BR for $y$ as a function of $x$}
        \label{fig:fig:ChickenGame_greedy y_br}
    \end{subfigure}
    \hfill
    \begin{subfigure}[t]{0.45\textwidth}
        \centering
        \includegraphics[width=\textwidth]{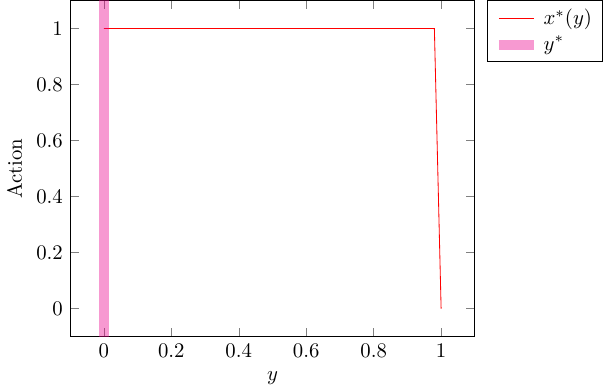}
        \caption{BR for $x$ as a function of $y$}
        \label{fig:fig:ChickenGame_greedy x_br}
    \end{subfigure}
    \newline
    \begin{subfigure}[t]{0.45\textwidth}
        \centering
        \includegraphics[width=\textwidth]{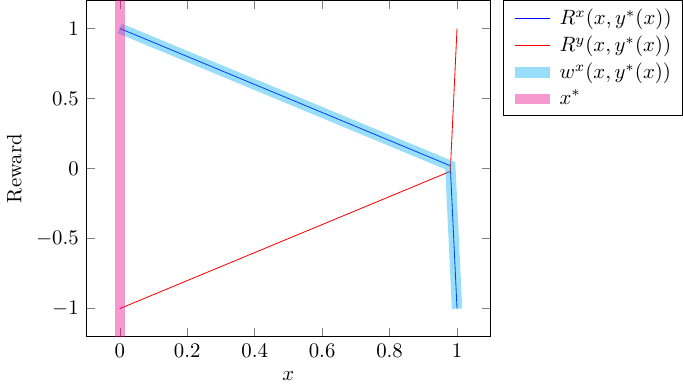}
        \caption{Rewards as a function of $x$}
        \label{fig:fig:ChickenGame_greedy br_rewards_x}
    \end{subfigure}
    \hfill
    \begin{subfigure}[t]{0.45\textwidth}
        \centering
        \includegraphics[width=\textwidth]{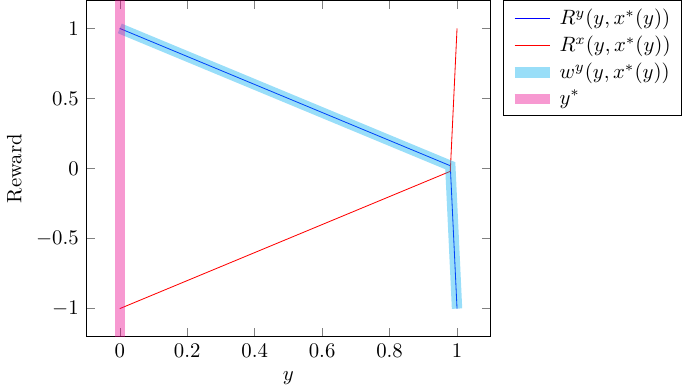}
        \caption{Rewards as a function of $y$}
        \label{fig:fig:ChickenGame_greedy br_rewards_y}
    \end{subfigure}
    \caption{Stackelberg strategy profile (Greedy WE) for ChickenGame \\ $x^* = 0.000, y^* = 0.000, R^x = -100.000, R^y = -100.000$}
    \label{fig:ChickenGame_greedy}
\end{figure*}

\begin{figure*}
    \centering
    \captionsetup{justification=centering}
    \begin{subfigure}[t]{0.45\textwidth}
        \centering
        \includegraphics[width=\textwidth]{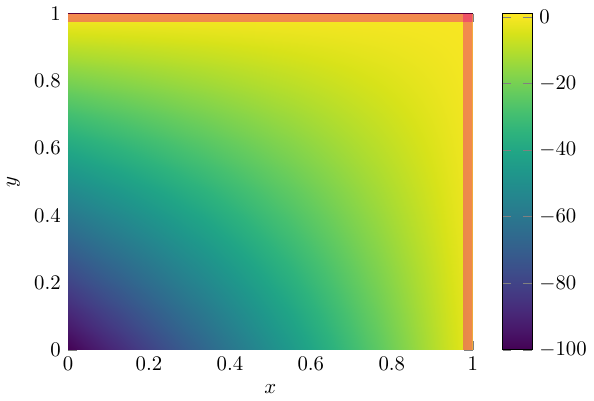}
        \caption{Reward surface for $x$}
        \label{fig:fig:ChickenGame_egalitarian reward_surface_x}
    \end{subfigure}
    \hfill
    \begin{subfigure}[t]{0.45\textwidth}
        \centering
        \includegraphics[width=\textwidth]{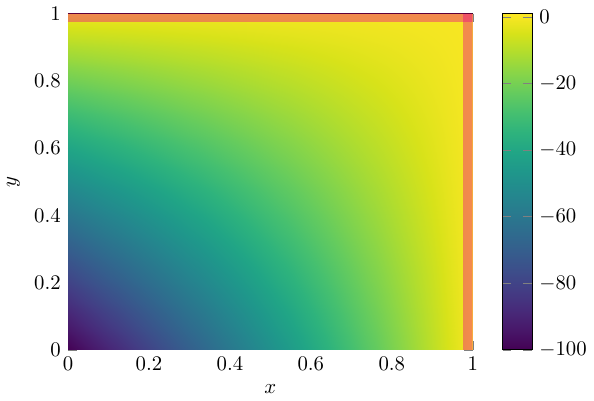}
        \caption{Reward surface for $y$}
        \label{fig:fig:ChickenGame_egalitarian reward_surface_y}
    \end{subfigure}
    \newline
    \begin{subfigure}[t]{0.45\textwidth}
        \centering
        \includegraphics[width=\textwidth]{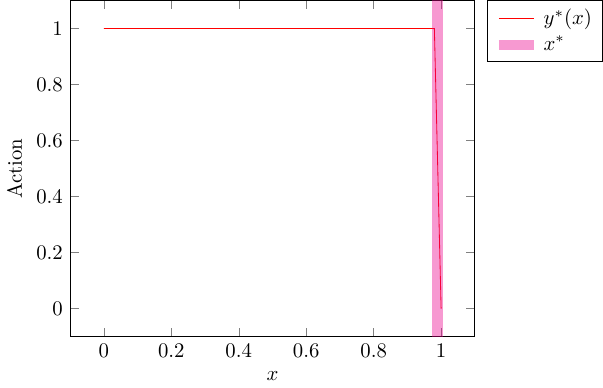}
        \caption{BR for $y$ as a function of $x$}
        \label{fig:fig:ChickenGame_egalitarian y_br}
    \end{subfigure}
    \hfill
    \begin{subfigure}[t]{0.45\textwidth}
        \centering
        \includegraphics[width=\textwidth]{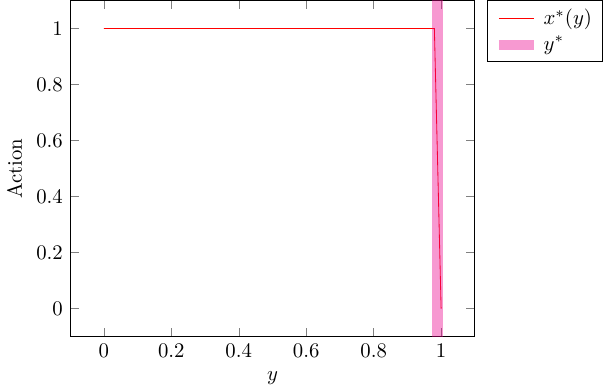}
        \caption{BR for $x$ as a function of $y$}
        \label{fig:fig:ChickenGame_egalitarian x_br}
    \end{subfigure}
    \newline
    \begin{subfigure}[t]{0.45\textwidth}
        \centering
        \includegraphics[width=\textwidth]{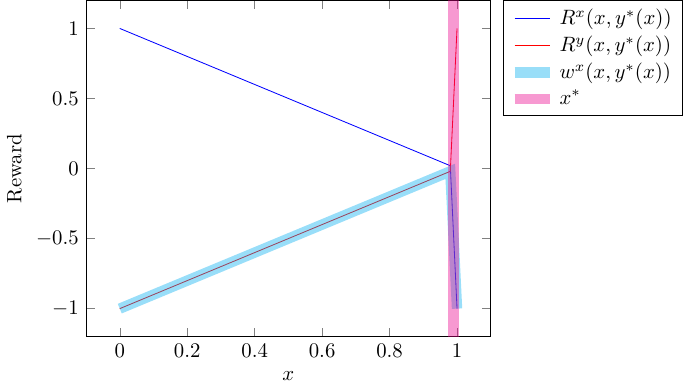}
        \caption{Rewards as a function of $x$}
        \label{fig:fig:ChickenGame_egalitarian br_rewards_x}
    \end{subfigure}
    \hfill
    \begin{subfigure}[t]{0.45\textwidth}
        \centering
        \includegraphics[width=\textwidth]{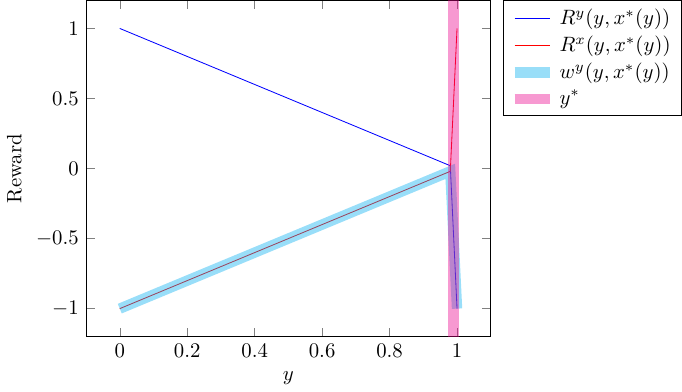}
        \caption{Rewards as a function of $y$}
        \label{fig:fig:ChickenGame_egalitarian br_rewards_y}
    \end{subfigure}
    \caption{Egalitarian WE for ChickenGame \\ $x^* = 0.989, y^* = 0.989, R^x = -0.011, R^y = -0.011$}
    \label{fig:ChickenGame_egalitarian}
\end{figure*}

\begin{figure*}
    \centering
    \captionsetup{justification=centering}
    \begin{subfigure}[t]{0.45\textwidth}
        \centering
        \includegraphics[width=\textwidth]{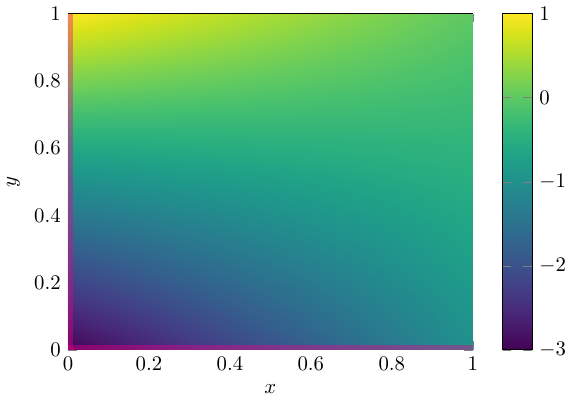}
        \caption{Reward surface for $x$}
        \label{fig:fig:BabyChickenGame_greedy reward_surface_x}
    \end{subfigure}
    \hfill
    \begin{subfigure}[t]{0.45\textwidth}
        \centering
        \includegraphics[width=\textwidth]{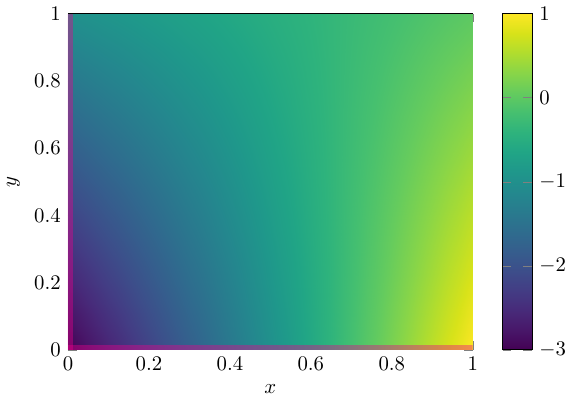}
        \caption{Reward surface for $y$}
        \label{fig:fig:BabyChickenGame_greedy reward_surface_y}
    \end{subfigure}
    \newline
    \begin{subfigure}[t]{0.45\textwidth}
        \centering
        \includegraphics[width=\textwidth]{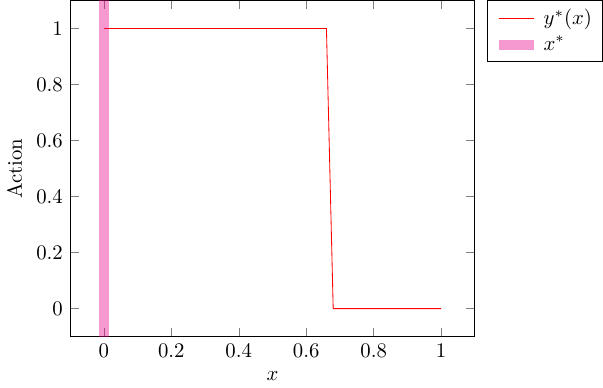}
        \caption{BR for $y$ as a function of $x$}
        \label{fig:fig:BabyChickenGame_greedy y_br}
    \end{subfigure}
    \hfill
    \begin{subfigure}[t]{0.45\textwidth}
        \centering
        \includegraphics[width=\textwidth]{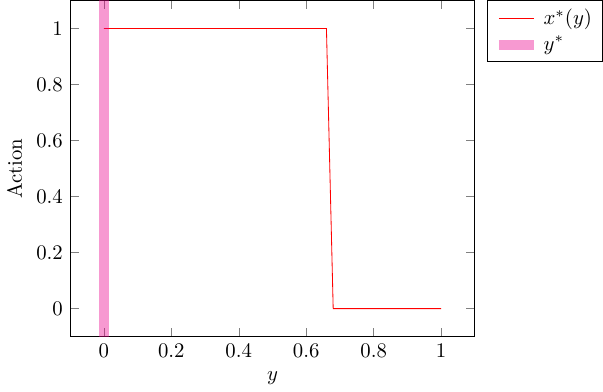}
        \caption{BR for $x$ as a function of $y$}
        \label{fig:fig:BabyChickenGame_greedy x_br}
    \end{subfigure}
    \newline
    \begin{subfigure}[t]{0.45\textwidth}
        \centering
        \includegraphics[width=\textwidth]{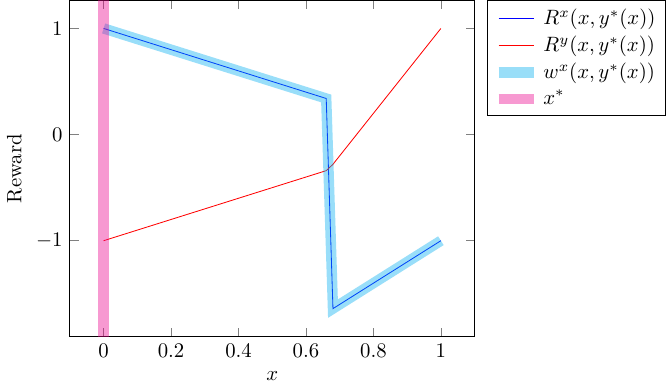}
        \caption{Rewards as a function of $x$}
        \label{fig:fig:BabyChickenGame_greedy br_rewards_x}
    \end{subfigure}
    \hfill
    \begin{subfigure}[t]{0.45\textwidth}
        \centering
        \includegraphics[width=\textwidth]{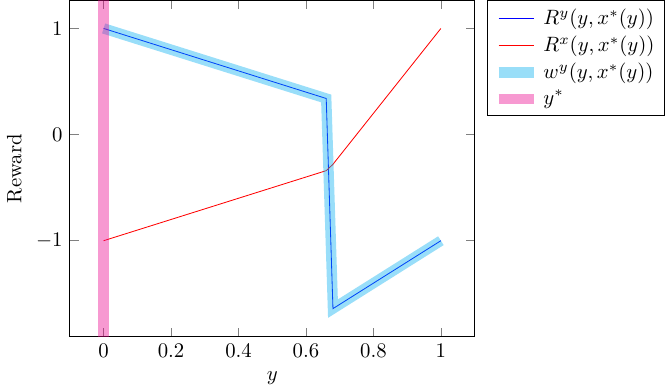}
        \caption{Rewards as a function of $y$}
        \label{fig:fig:BabyChickenGame_greedy br_rewards_y}
    \end{subfigure}
    \caption{Stackelberg strategy profile (Greedy WE) for BabyChickenGame \\ $x^* = 0.000, y^* = 0.000, R^x = -3.000, R^y = -3.000$}
    \label{fig:BabyChickenGame_greedy}
\end{figure*}

\begin{figure*}
    \centering
    \captionsetup{justification=centering}
    \begin{subfigure}[t]{0.45\textwidth}
        \centering
        \includegraphics[width=\textwidth]{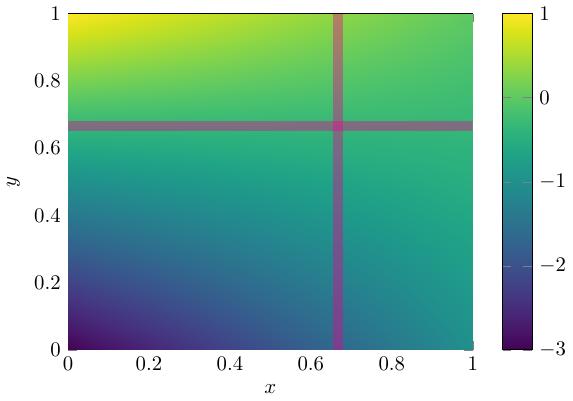}
        \caption{Reward surface for $x$}
        \label{fig:fig:BabyChickenGame_egalitarian reward_surface_x}
    \end{subfigure}
    \hfill
    \begin{subfigure}[t]{0.45\textwidth}
        \centering
        \includegraphics[width=\textwidth]{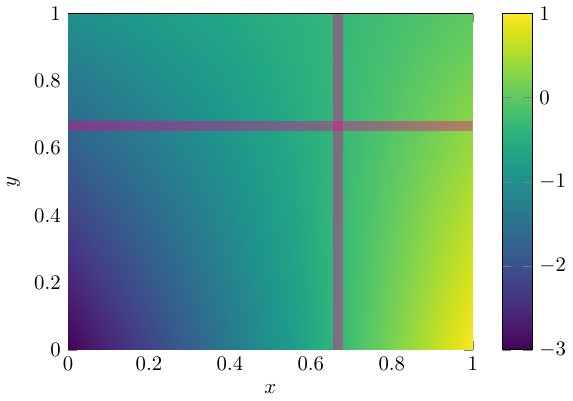}
        \caption{Reward surface for $y$}
        \label{fig:fig:BabyChickenGame_egalitarian reward_surface_y}
    \end{subfigure}
    \newline
    \begin{subfigure}[t]{0.45\textwidth}
        \centering
        \includegraphics[width=\textwidth]{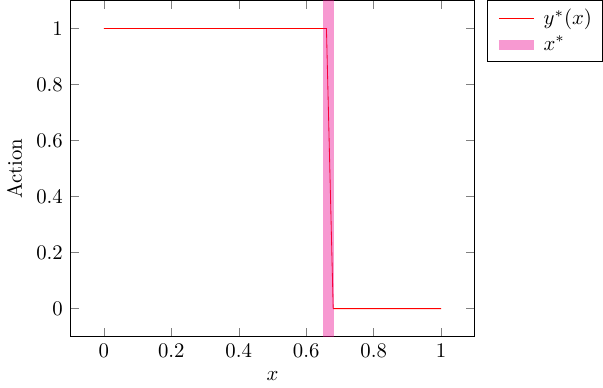}
        \caption{BR for $y$ as a function of $x$}
        \label{fig:fig:BabyChickenGame_egalitarian y_br}
    \end{subfigure}
    \hfill
    \begin{subfigure}[t]{0.45\textwidth}
        \centering
        \includegraphics[width=\textwidth]{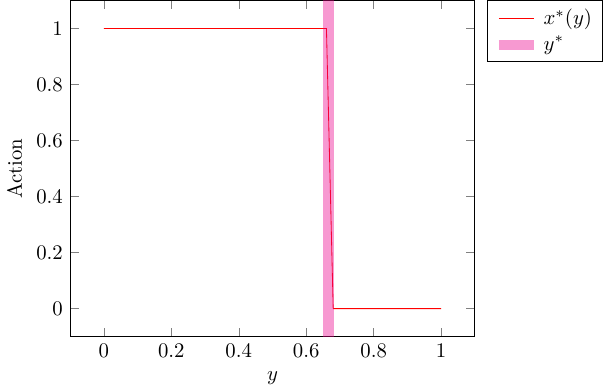}
        \caption{BR for $x$ as a function of $y$}
        \label{fig:fig:BabyChickenGame_egalitarian x_br}
    \end{subfigure}
    \newline
    \begin{subfigure}[t]{0.45\textwidth}
        \centering
        \includegraphics[width=\textwidth]{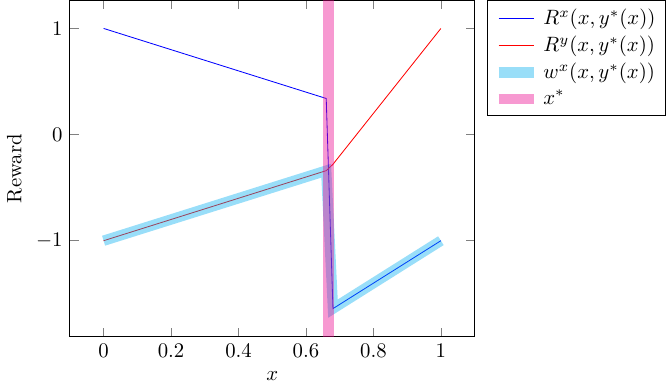}
        \caption{Rewards as a function of $x$}
        \label{fig:fig:BabyChickenGame_egalitarian br_rewards_x}
    \end{subfigure}
    \hfill
    \begin{subfigure}[t]{0.45\textwidth}
        \centering
        \includegraphics[width=\textwidth]{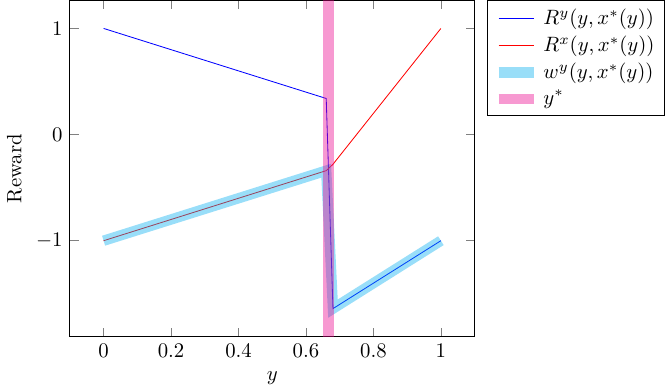}
        \caption{Rewards as a function of $y$}
        \label{fig:fig:BabyChickenGame_egalitarian br_rewards_y}
    \end{subfigure}
    \caption{Egalitarian WE for BabyChickenGame \\ $x^* = 0.666, y^* = 0.666, R^x = -0.334, R^y = -0.334$}
    \label{fig:BabyChickenGame_egalitarian}
\end{figure*}

\begin{figure*}
    \centering
    \captionsetup{justification=centering}
    \begin{subfigure}[t]{0.45\textwidth}
        \centering
        \includegraphics[width=\textwidth]{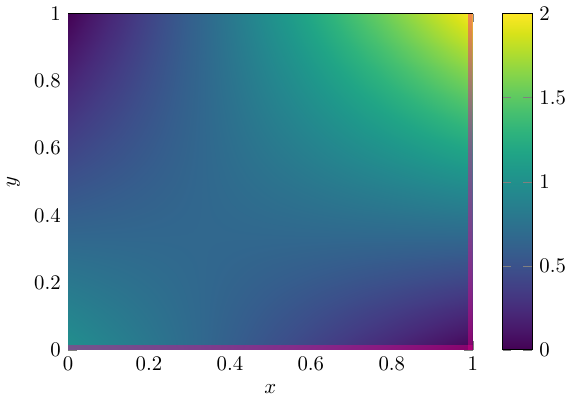}
        \caption{Reward surface for $x$}
        \label{fig:fig:CoordinationGame_greedy reward_surface_x}
    \end{subfigure}
    \hfill
    \begin{subfigure}[t]{0.45\textwidth}
        \centering
        \includegraphics[width=\textwidth]{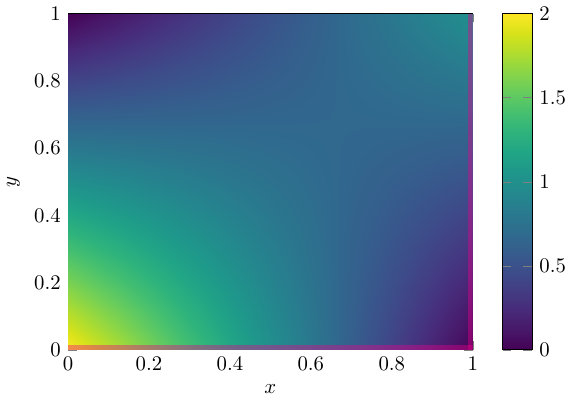}
        \caption{Reward surface for $y$}
        \label{fig:fig:CoordinationGame_greedy reward_surface_y}
    \end{subfigure}
    \newline
    \begin{subfigure}[t]{0.45\textwidth}
        \centering
        \includegraphics[width=\textwidth]{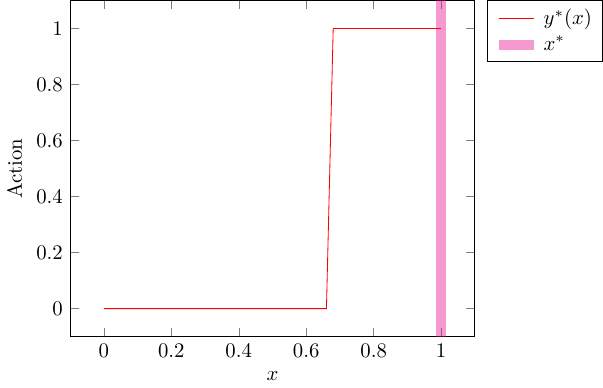}
        \caption{BR for $y$ as a function of $x$}
        \label{fig:fig:CoordinationGame_greedy y_br}
    \end{subfigure}
    \hfill
    \begin{subfigure}[t]{0.45\textwidth}
        \centering
        \includegraphics[width=\textwidth]{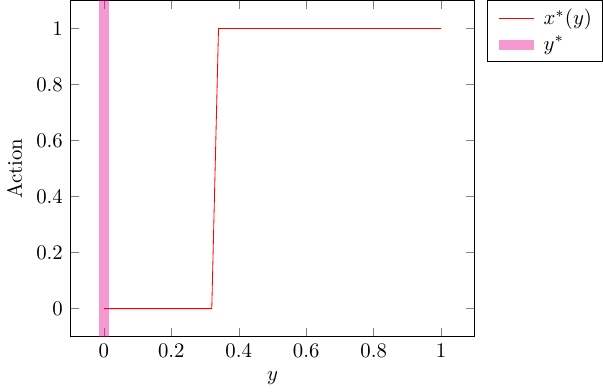}
        \caption{BR for $x$ as a function of $y$}
        \label{fig:fig:CoordinationGame_greedy x_br}
    \end{subfigure}
    \newline
    \begin{subfigure}[t]{0.45\textwidth}
        \centering
        \includegraphics[width=\textwidth]{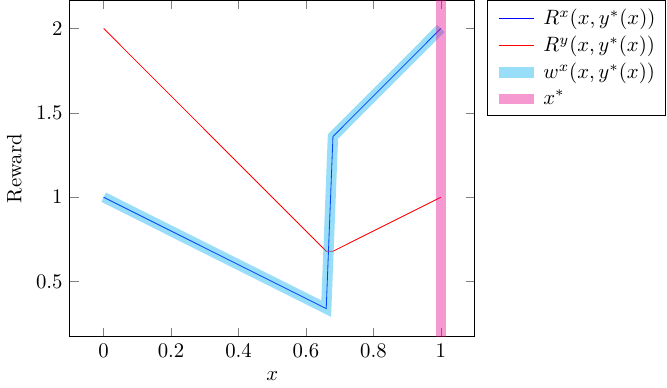}
        \caption{Rewards as a function of $x$}
        \label{fig:fig:CoordinationGame_greedy br_rewards_x}
    \end{subfigure}
    \hfill
    \begin{subfigure}[t]{0.45\textwidth}
        \centering
        \includegraphics[width=\textwidth]{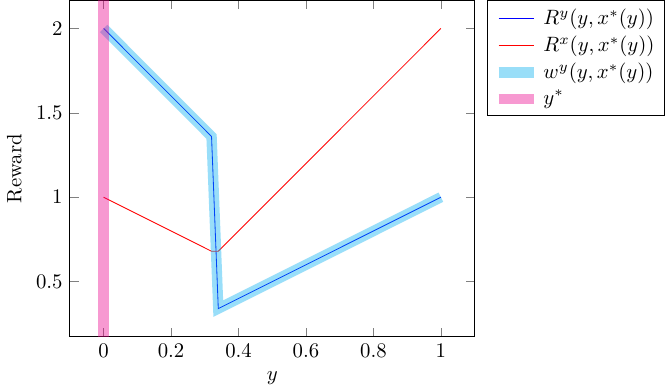}
        \caption{Rewards as a function of $y$}
        \label{fig:fig:CoordinationGame_greedy br_rewards_y}
    \end{subfigure}
    \caption{Stackelberg strategy profile (Greedy WE) for Bach Or Stravinsky \\ $x^* = 1.000, y^* = 0.000, R^x = 0.000, R^y = 0.000$}
    \label{fig:CoordinationGame_greedy}
\end{figure*}

\begin{figure*}
    \centering
    \captionsetup{justification=centering}
    \begin{subfigure}[t]{0.45\textwidth}
        \centering
        \includegraphics[width=\textwidth]{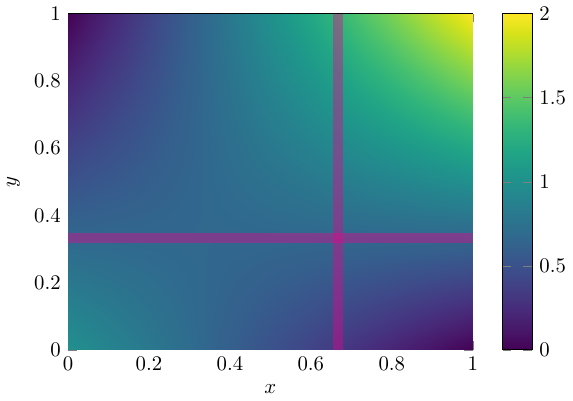}
        \caption{Reward surface for $x$}
        \label{fig:fig:CoordinationGame_fairness reward_surface_x}
    \end{subfigure}
    \hfill
    \begin{subfigure}[t]{0.45\textwidth}
        \centering
        \includegraphics[width=\textwidth]{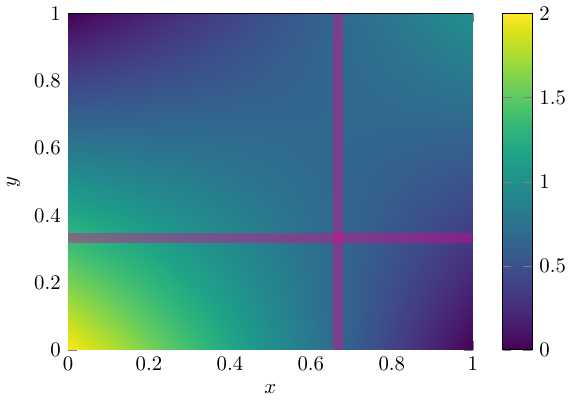}
        \caption{Reward surface for $y$}
        \label{fig:fig:CoordinationGame_fairness reward_surface_y}
    \end{subfigure}
    \newline
    \begin{subfigure}[t]{0.45\textwidth}
        \centering
        \includegraphics[width=\textwidth]{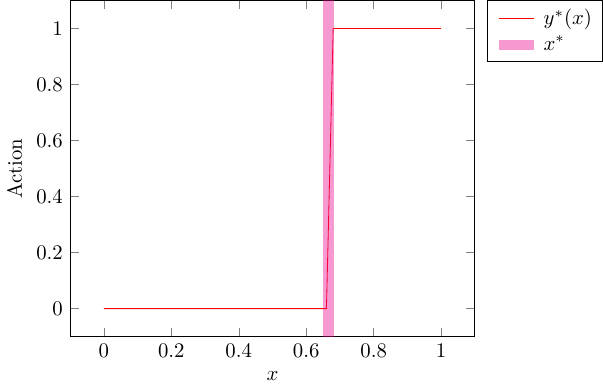}
        \caption{BR for $y$ as a function of $x$}
        \label{fig:fig:CoordinationGame_fairness y_br}
    \end{subfigure}
    \hfill
    \begin{subfigure}[t]{0.45\textwidth}
        \centering
        \includegraphics[width=\textwidth]{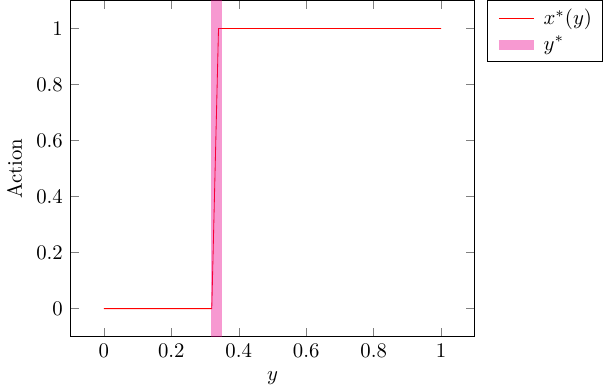}
        \caption{BR for $x$ as a function of $y$}
        \label{fig:fig:CoordinationGame_fairness x_br}
    \end{subfigure}
    \newline
    \begin{subfigure}[t]{0.45\textwidth}
        \centering
        \includegraphics[width=\textwidth]{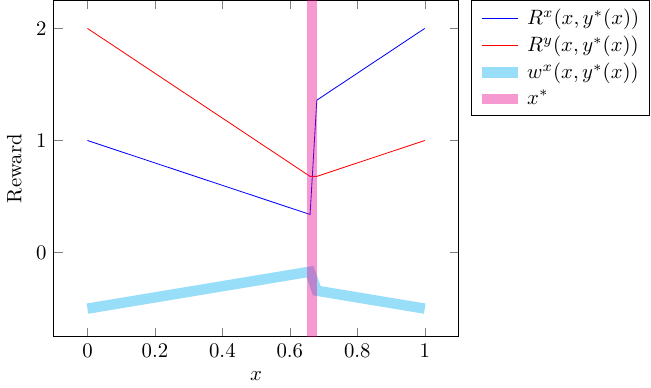}
        \caption{Rewards as a function of $x$}
        \label{fig:fig:CoordinationGame_fairness br_rewards_x}
    \end{subfigure}
    \hfill
    \begin{subfigure}[t]{0.45\textwidth}
        \centering
        \includegraphics[width=\textwidth]{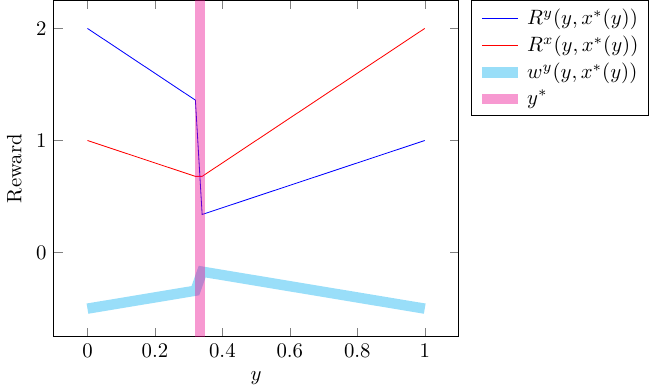}
        \caption{Rewards as a function of $y$}
        \label{fig:fig:CoordinationGame_fairness br_rewards_y}
    \end{subfigure}
    \caption{Fairness WE for Bach Or Stravinsky \\ $x^* = 0.666, y^* = 0.334, R^x = 0.667, R^y = 0.667$}
    \label{fig:CoordinationGame_fairness}
\end{figure*}

\begin{figure*}
    \centering
    \captionsetup{justification=centering}
    \begin{subfigure}[t]{0.45\textwidth}
        \centering
        \includegraphics[width=\textwidth]{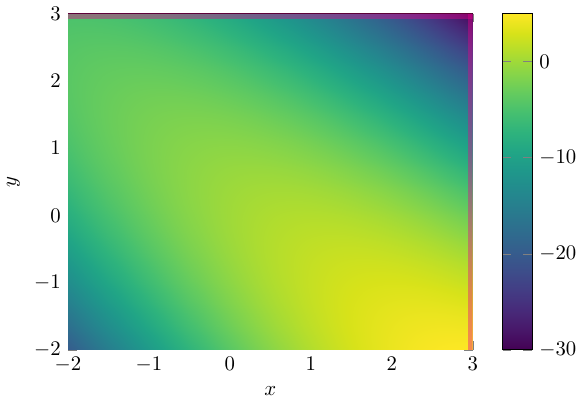}
        \caption{Reward surface for $x$}
        \label{fig:fig:Tandem_greedy reward_surface_x}
    \end{subfigure}
    \hfill
    \begin{subfigure}[t]{0.45\textwidth}
        \centering
        \includegraphics[width=\textwidth]{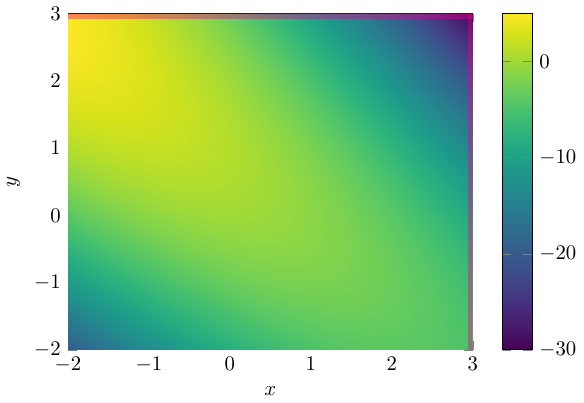}
        \caption{Reward surface for $y$}
        \label{fig:fig:Tandem_greedy reward_surface_y}
    \end{subfigure}
    \newline
    \begin{subfigure}[t]{0.45\textwidth}
        \centering
        \includegraphics[width=\textwidth]{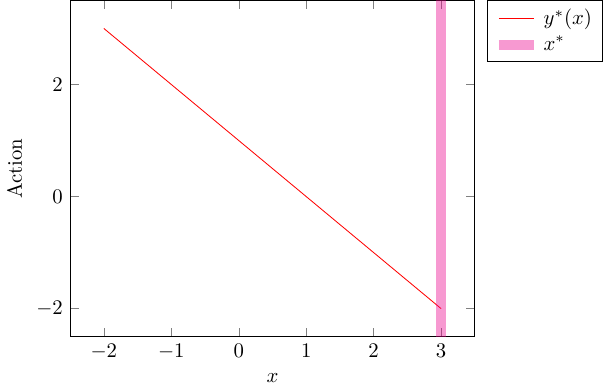}
        \caption{BR for $y$ as a function of $x$}
        \label{fig:fig:Tandem_greedy y_br}
    \end{subfigure}
    \hfill
    \begin{subfigure}[t]{0.45\textwidth}
        \centering
        \includegraphics[width=\textwidth]{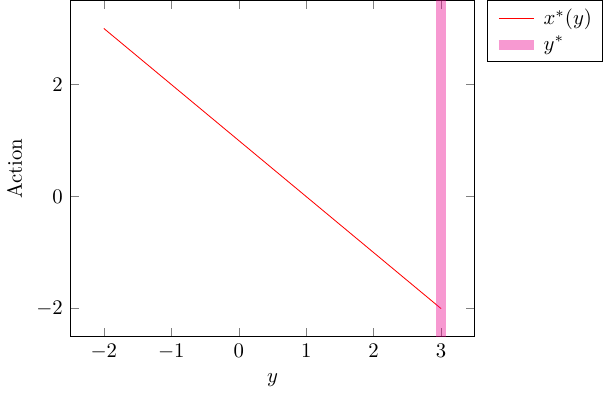}
        \caption{BR for $x$ as a function of $y$}
        \label{fig:fig:Tandem_greedy x_br}
    \end{subfigure}
    \newline
    \begin{subfigure}[t]{0.45\textwidth}
        \centering
        \includegraphics[width=\textwidth]{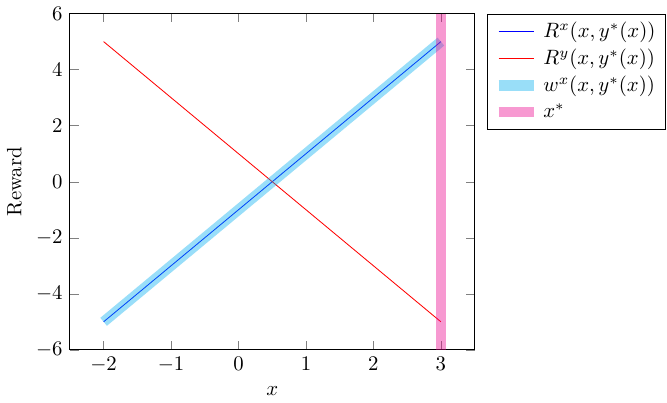}
        \caption{Rewards as a function of $x$}
        \label{fig:fig:Tandem_greedy br_rewards_x}
    \end{subfigure}
    \hfill
    \begin{subfigure}[t]{0.45\textwidth}
        \centering
        \includegraphics[width=\textwidth]{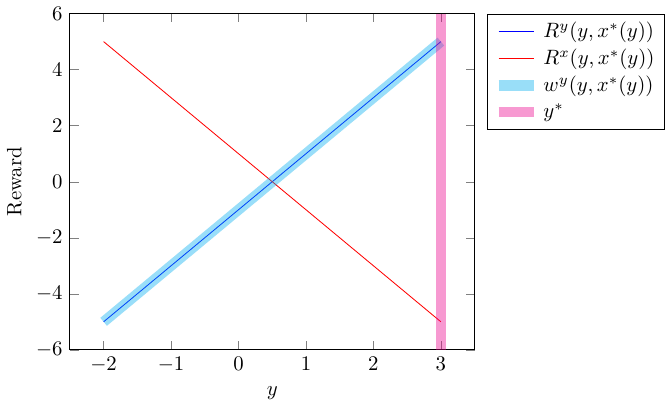}
        \caption{Rewards as a function of $y$}
        \label{fig:fig:Tandem_greedy br_rewards_y}
    \end{subfigure}
    \caption{Stackelberg strategy profile (Greedy WE) for Tandem \\ $x^* = 3.000, y^* = 3.000, R^x = -30.000, R^y = -30.000$}
    \label{fig:Tandem_greedy}
\end{figure*}

\begin{figure*}
    \centering
    \captionsetup{justification=centering}
    \begin{subfigure}[t]{0.45\textwidth}
        \centering
        \includegraphics[width=\textwidth]{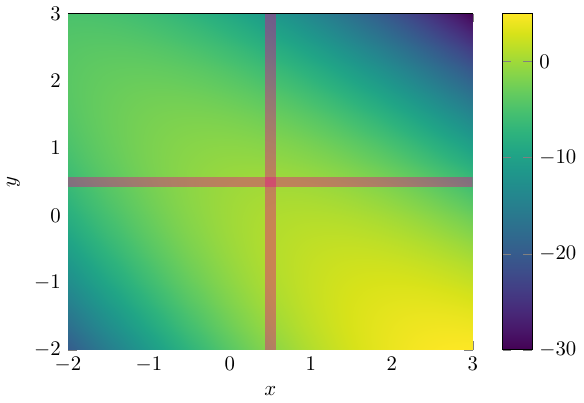}
        \caption{Reward surface for $x$}
        \label{fig:fig:Tandem_egalitarian reward_surface_x}
    \end{subfigure}
    \hfill
    \begin{subfigure}[t]{0.45\textwidth}
        \centering
        \includegraphics[width=\textwidth]{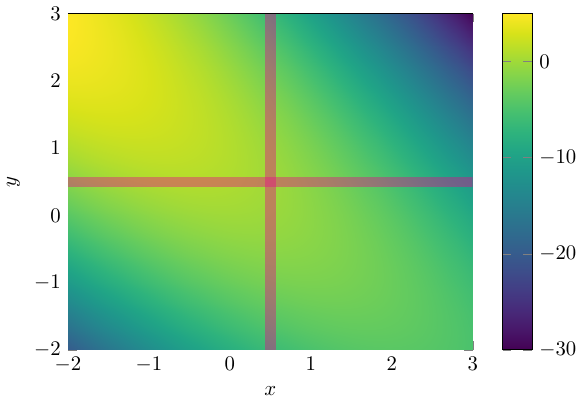}
        \caption{Reward surface for $y$}
        \label{fig:fig:Tandem_egalitarian reward_surface_y}
    \end{subfigure}
    \newline
    \begin{subfigure}[t]{0.45\textwidth}
        \centering
        \includegraphics[width=\textwidth]{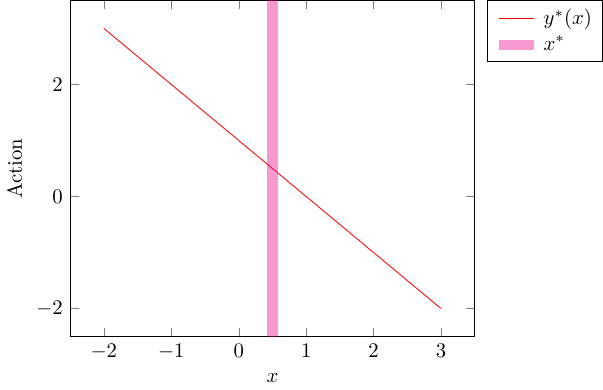}
        \caption{BR for $y$ as a function of $x$}
        \label{fig:fig:Tandem_egalitarian y_br}
    \end{subfigure}
    \hfill
    \begin{subfigure}[t]{0.45\textwidth}
        \centering
        \includegraphics[width=\textwidth]{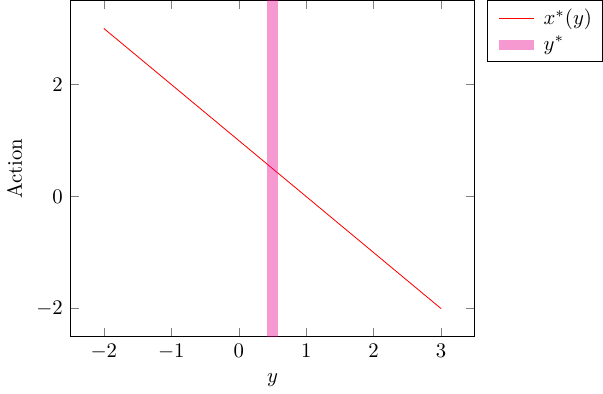}
        \caption{BR for $x$ as a function of $y$}
        \label{fig:fig:Tandem_egalitarian x_br}
    \end{subfigure}
    \newline
    \begin{subfigure}[t]{0.45\textwidth}
        \centering
        \includegraphics[width=\textwidth]{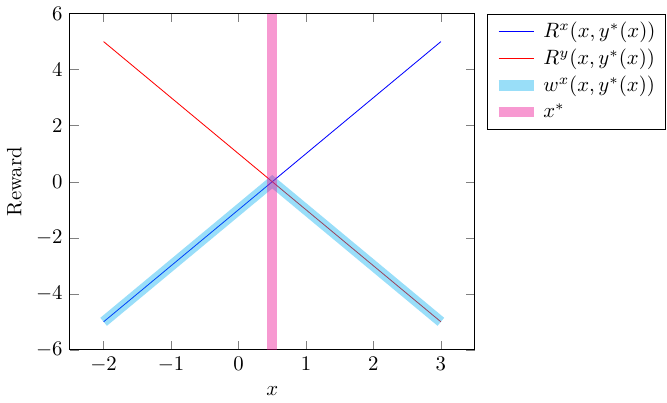}
        \caption{Rewards as a function of $x$}
        \label{fig:fig:Tandem_egalitarian br_rewards_x}
    \end{subfigure}
    \hfill
    \begin{subfigure}[t]{0.45\textwidth}
        \centering
        \includegraphics[width=\textwidth]{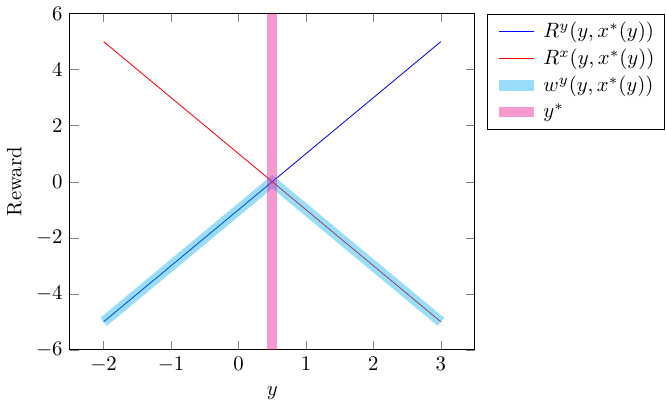}
        \caption{Rewards as a function of $y$}
        \label{fig:fig:Tandem_egalitarian br_rewards_y}
    \end{subfigure}
    \caption{Egalitarian WE for Tandem \\ $x^* = 0.500, y^* = 0.500, R^x = 0.000, R^y = 0.000$}
    \label{fig:Tandem_egalitarian}
\end{figure*}

\begin{figure*}
    \centering
    \captionsetup{justification=centering}
    \begin{subfigure}[t]{0.45\textwidth}
        \centering
        \includegraphics[width=\textwidth]{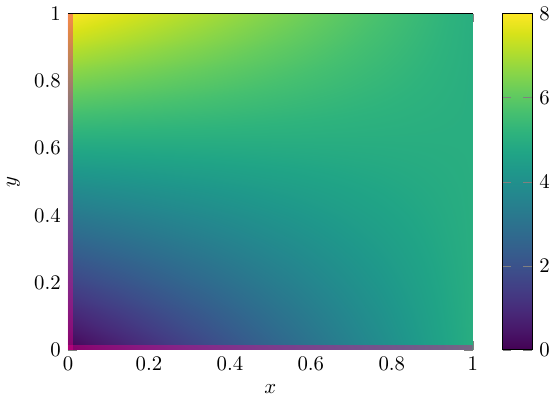}
        \caption{Reward surface for $x$}
        \label{fig:fig:UltimatumGame_greedy reward_surface_x}
    \end{subfigure}
    \hfill
    \begin{subfigure}[t]{0.45\textwidth}
        \centering
        \includegraphics[width=\textwidth]{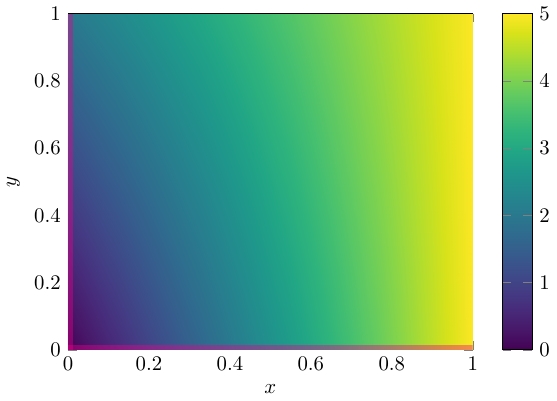}
        \caption{Reward surface for $y$}
        \label{fig:fig:UltimatumGame_greedy reward_surface_y}
    \end{subfigure}
    \newline
    \begin{subfigure}[t]{0.45\textwidth}
        \centering
        \includegraphics[width=\textwidth]{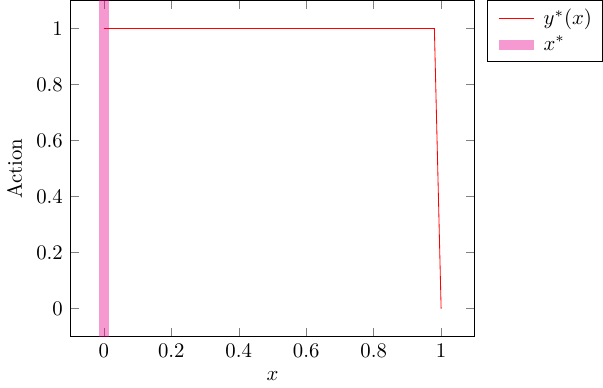}
        \caption{BR for $y$ as a function of $x$}
        \label{fig:fig:UltimatumGame_greedy y_br}
    \end{subfigure}
    \hfill
    \begin{subfigure}[t]{0.45\textwidth}
        \centering
        \includegraphics[width=\textwidth]{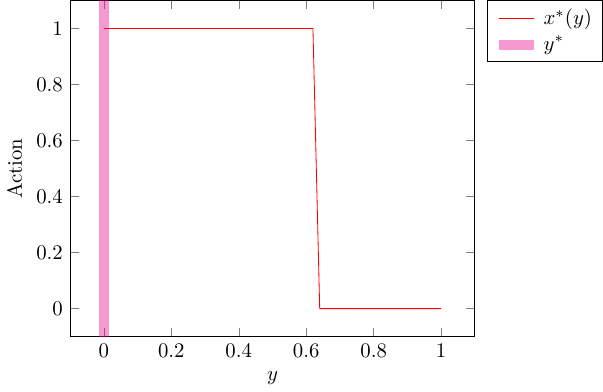}
        \caption{BR for $x$ as a function of $y$}
        \label{fig:fig:UltimatumGame_greedy x_br}
    \end{subfigure}
    \newline
    \begin{subfigure}[t]{0.45\textwidth}
        \centering
        \includegraphics[width=\textwidth]{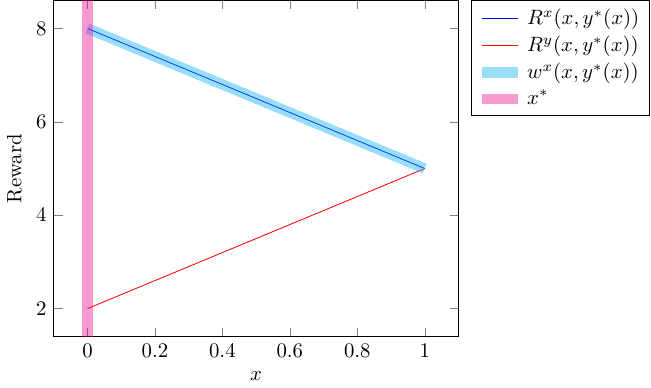}
        \caption{Rewards as a function of $x$}
        \label{fig:fig:UltimatumGame_greedy br_rewards_x}
    \end{subfigure}
    \hfill
    \begin{subfigure}[t]{0.45\textwidth}
        \centering
        \includegraphics[width=\textwidth]{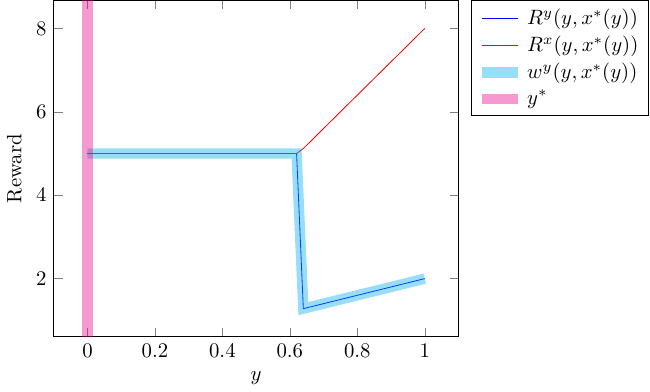}
        \caption{Rewards as a function of $y$}
        \label{fig:fig:UltimatumGame_greedy br_rewards_y}
    \end{subfigure}
    \caption{Stackelberg strategy profile (Greedy WE) for UltimatumGame \\ $x^* = 0.000, y^* = 0.000, R^x = 0.000, R^y = 0.000$}
    \label{fig:UltimatumGame_greedy}
\end{figure*}

\begin{figure*}
    \centering
    \captionsetup{justification=centering}
    \begin{subfigure}[t]{0.45\textwidth}
        \centering
        \includegraphics[width=\textwidth]{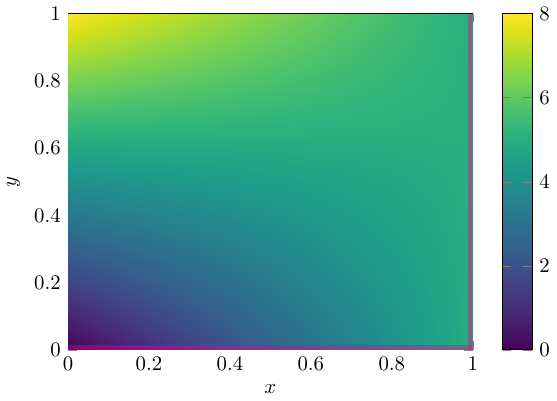}
        \caption{Reward surface for $x$}
        \label{fig:fig:UltimatumGame_egalitarian reward_surface_x}
    \end{subfigure}
    \hfill
    \begin{subfigure}[t]{0.45\textwidth}
        \centering
        \includegraphics[width=\textwidth]{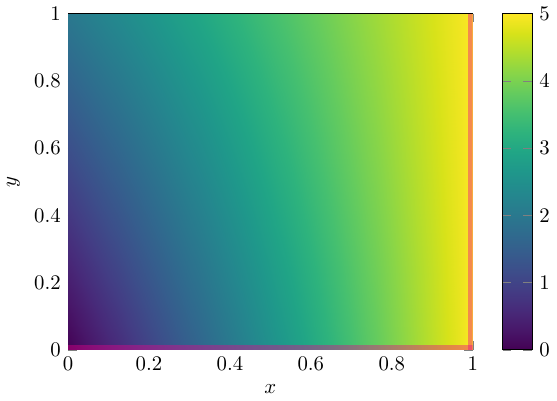}
        \caption{Reward surface for $y$}
        \label{fig:fig:UltimatumGame_egalitarian reward_surface_y}
    \end{subfigure}
    \newline
    \begin{subfigure}[t]{0.45\textwidth}
        \centering
        \includegraphics[width=\textwidth]{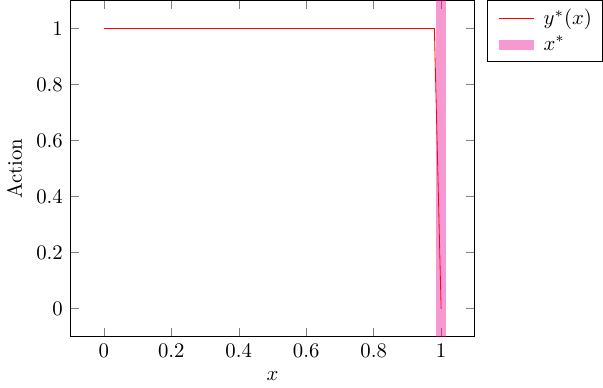}
        \caption{BR for $y$ as a function of $x$}
        \label{fig:fig:UltimatumGame_egalitarian y_br}
    \end{subfigure}
    \hfill
    \begin{subfigure}[t]{0.45\textwidth}
        \centering
        \includegraphics[width=\textwidth]{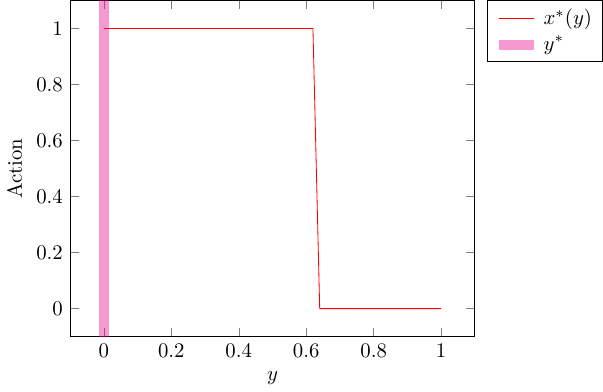}
        \caption{BR for $x$ as a function of $y$}
        \label{fig:fig:UltimatumGame_egalitarian x_br}
    \end{subfigure}
    \newline
    \begin{subfigure}[t]{0.45\textwidth}
        \centering
        \includegraphics[width=\textwidth]{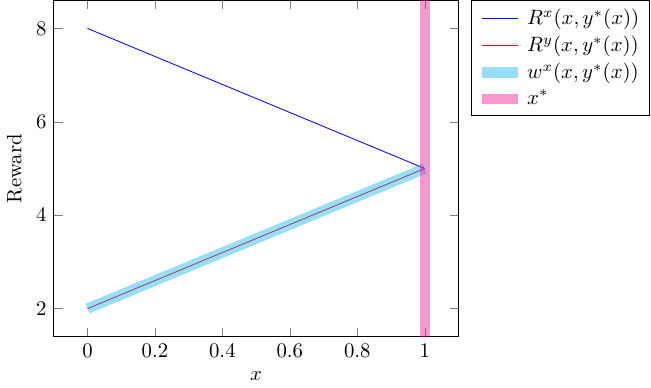}
        \caption{Rewards as a function of $x$}
        \label{fig:fig:UltimatumGame_egalitarian br_rewards_x}
    \end{subfigure}
    \hfill
    \begin{subfigure}[t]{0.45\textwidth}
        \centering
        \includegraphics[width=\textwidth]{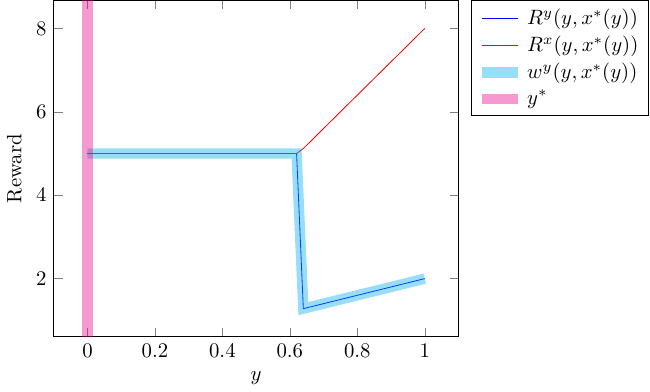}
        \caption{Rewards as a function of $y$}
        \label{fig:fig:UltimatumGame_egalitarian br_rewards_y}
    \end{subfigure}
    \caption{Egalitarian WE for UltimatumGame \\ $x^* = 1.000, y^* = 0.000, R^x = 5.000, R^y = 5.000$}
    \label{fig:UltimatumGame_egalitarian}
\end{figure*}

\begin{figure*}
    \centering
    \captionsetup{justification=centering}
    \begin{subfigure}[t]{0.45\textwidth}
        \centering
        \includegraphics[width=\textwidth]{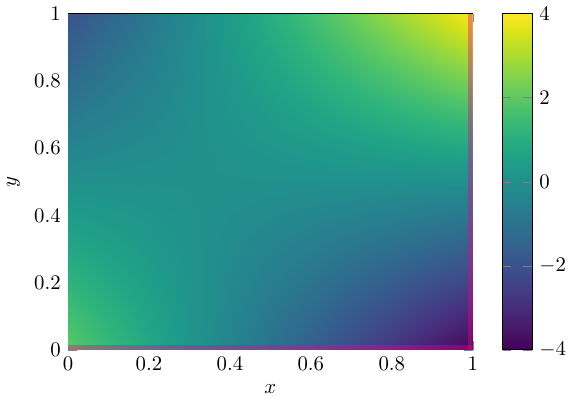}
        \caption{Reward surface for $x$}
        \label{fig:fig:EagleGame_greedy reward_surface_x}
    \end{subfigure}
    \hfill
    \begin{subfigure}[t]{0.45\textwidth}
        \centering
        \includegraphics[width=\textwidth]{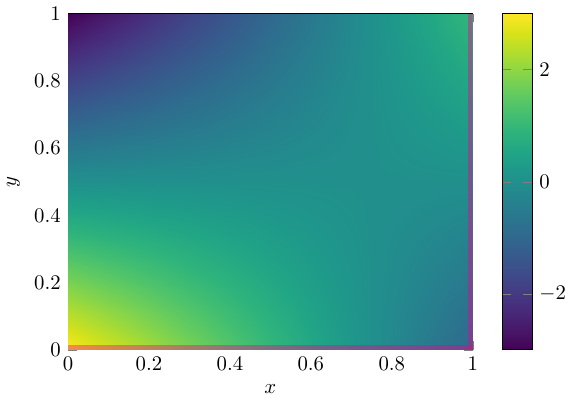}
        \caption{Reward surface for $y$}
        \label{fig:fig:EagleGame_greedy reward_surface_y}
    \end{subfigure}
    \newline
    \begin{subfigure}[t]{0.45\textwidth}
        \centering
        \includegraphics[width=\textwidth]{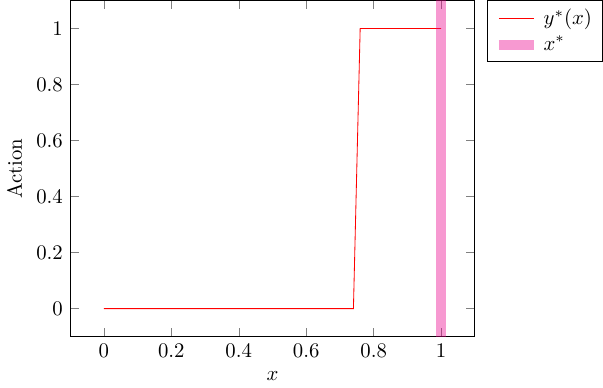}
        \caption{BR for $y$ as a function of $x$}
        \label{fig:fig:EagleGame_greedy y_br}
    \end{subfigure}
    \hfill
    \begin{subfigure}[t]{0.45\textwidth}
        \centering
        \includegraphics[width=\textwidth]{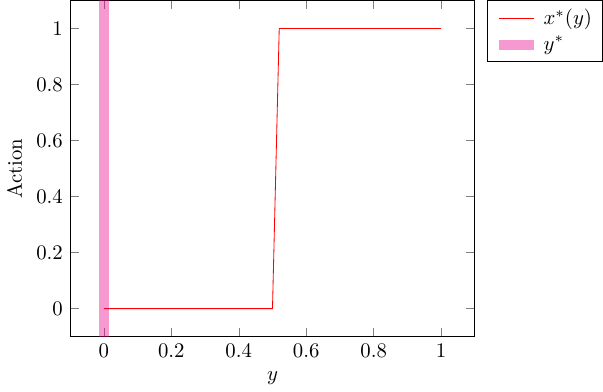}
        \caption{BR for $x$ as a function of $y$}
        \label{fig:fig:EagleGame_greedy x_br}
    \end{subfigure}
    \newline
    \begin{subfigure}[t]{0.45\textwidth}
        \centering
        \includegraphics[width=\textwidth]{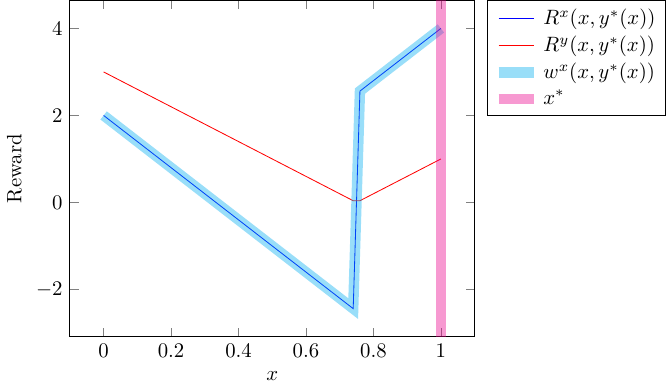}
        \caption{Rewards as a function of $x$}
        \label{fig:fig:EagleGame_greedy br_rewards_x}
    \end{subfigure}
    \hfill
    \begin{subfigure}[t]{0.45\textwidth}
        \centering
        \includegraphics[width=\textwidth]{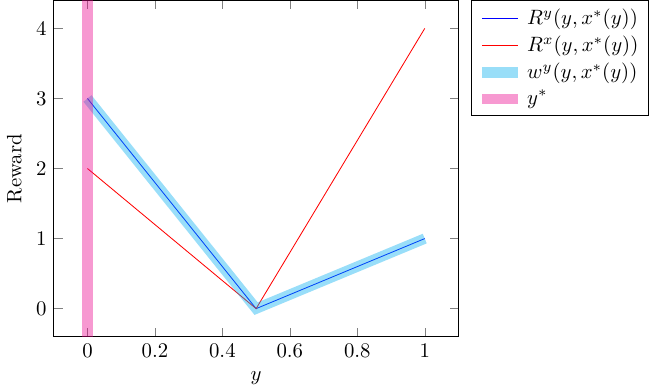}
        \caption{Rewards as a function of $y$}
        \label{fig:fig:EagleGame_greedy br_rewards_y}
    \end{subfigure}
    \caption{Stackelberg strategy profile (Greedy WE) for EagleGame \\ $x^* = 1.000, y^* = 0.000, R^x = -4.000, R^y = -1.000$}
    \label{fig:EagleGame_greedy}
\end{figure*}

\begin{figure*}
    \centering
    \captionsetup{justification=centering}
    \begin{subfigure}[t]{0.45\textwidth}
        \centering
        \includegraphics[width=\textwidth]{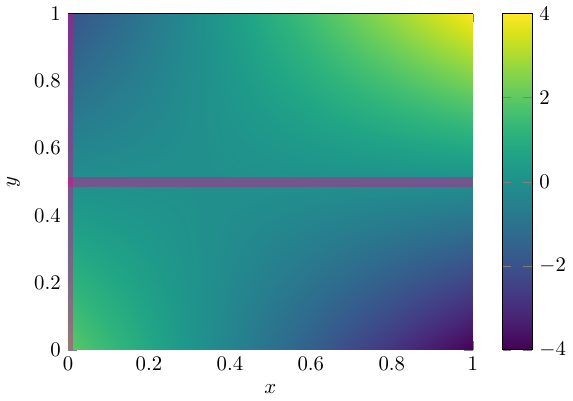}
        \caption{Reward surface for $x$}
        \label{fig:fig:EagleGame_fairness reward_surface_x}
    \end{subfigure}
    \hfill
    \begin{subfigure}[t]{0.45\textwidth}
        \centering
        \includegraphics[width=\textwidth]{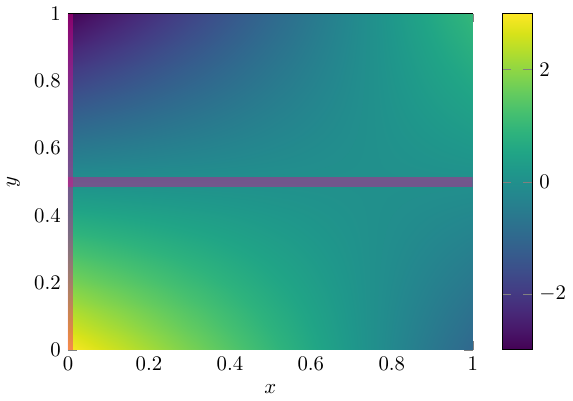}
        \caption{Reward surface for $y$}
        \label{fig:fig:EagleGame_fairness reward_surface_y}
    \end{subfigure}
    \newline
    \begin{subfigure}[t]{0.45\textwidth}
        \centering
        \includegraphics[width=\textwidth]{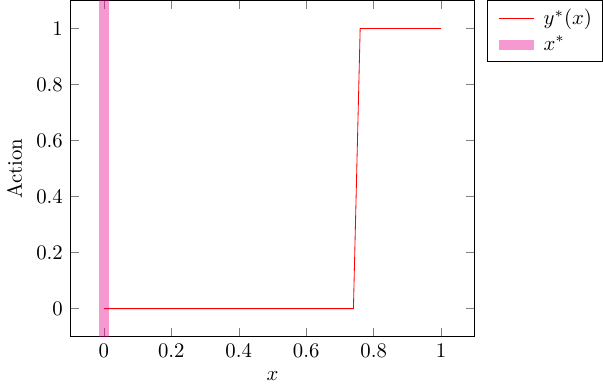}
        \caption{BR for $y$ as a function of $x$}
        \label{fig:fig:EagleGame_fairness y_br}
    \end{subfigure}
    \hfill
    \begin{subfigure}[t]{0.45\textwidth}
        \centering
        \includegraphics[width=\textwidth]{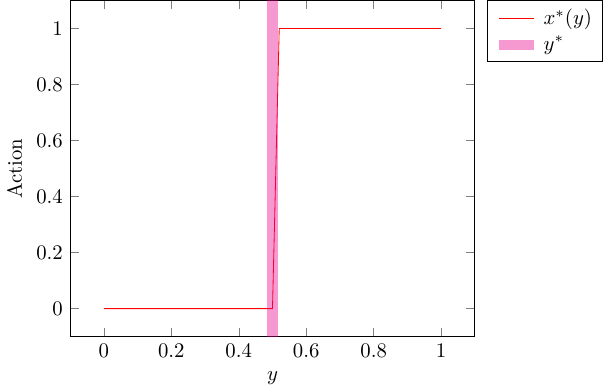}
        \caption{BR for $x$ as a function of $y$}
        \label{fig:fig:EagleGame_fairness x_br}
    \end{subfigure}
    \newline
    \begin{subfigure}[t]{0.45\textwidth}
        \centering
        \includegraphics[width=\textwidth]{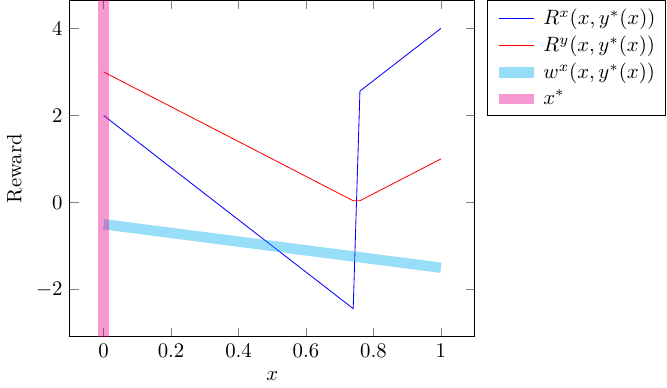}
        \caption{Rewards as a function of $x$}
        \label{fig:fig:EagleGame_fairness br_rewards_x}
    \end{subfigure}
    \hfill
    \begin{subfigure}[t]{0.45\textwidth}
        \centering
        \includegraphics[width=\textwidth]{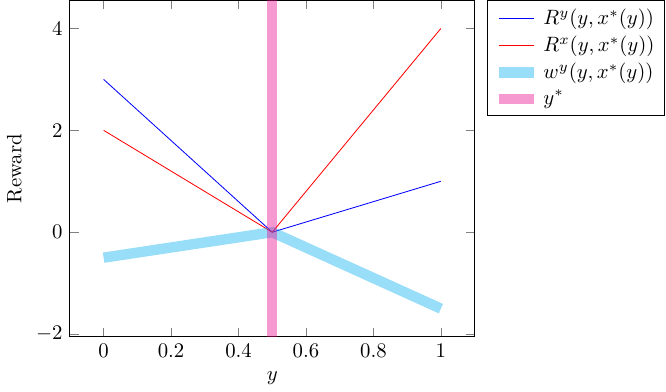}
        \caption{Rewards as a function of $y$}
        \label{fig:fig:EagleGame_fairness br_rewards_y}
    \end{subfigure}
    \caption{Fairness WE for EagleGame \\ $x^* = 0.000, y^* = 0.500, R^x = 0.000, R^y = 0.000$}
    \label{fig:EagleGame_fairness}
\end{figure*}

\begin{figure*}
    \centering
    \captionsetup{justification=centering}
    \begin{subfigure}[t]{0.45\textwidth}
        \centering
        \includegraphics[width=\textwidth]{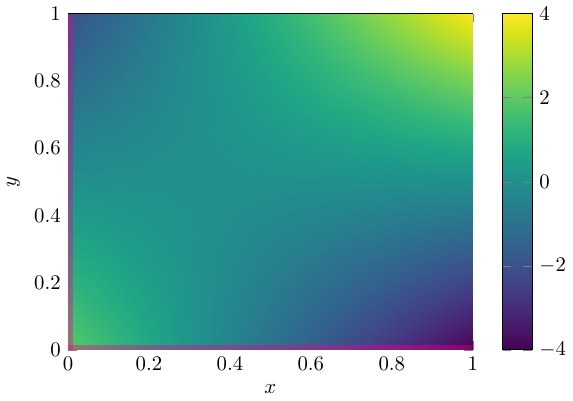}
        \caption{Reward surface for $x$}
        \label{fig:fig:EagleGame_egalitarian reward_surface_x}
    \end{subfigure}
    \hfill
    \begin{subfigure}[t]{0.45\textwidth}
        \centering
        \includegraphics[width=\textwidth]{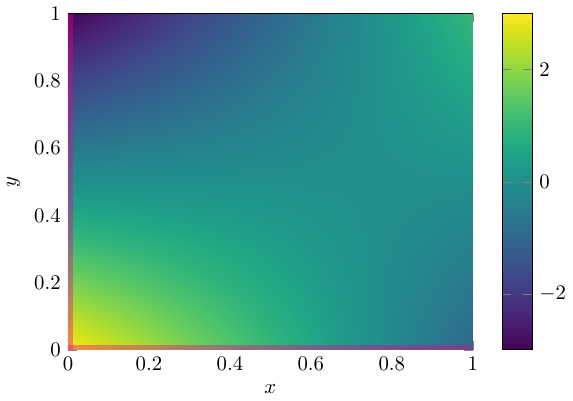}
        \caption{Reward surface for $y$}
        \label{fig:fig:EagleGame_egalitarian reward_surface_y}
    \end{subfigure}
    \newline
    \begin{subfigure}[t]{0.45\textwidth}
        \centering
        \includegraphics[width=\textwidth]{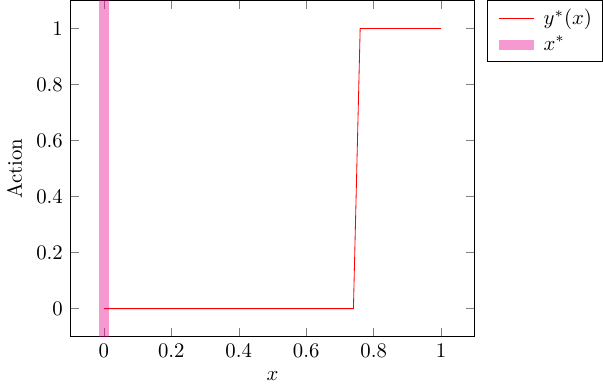}
        \caption{BR for $y$ as a function of $x$}
        \label{fig:fig:EagleGame_egalitarian y_br}
    \end{subfigure}
    \hfill
    \begin{subfigure}[t]{0.45\textwidth}
        \centering
        \includegraphics[width=\textwidth]{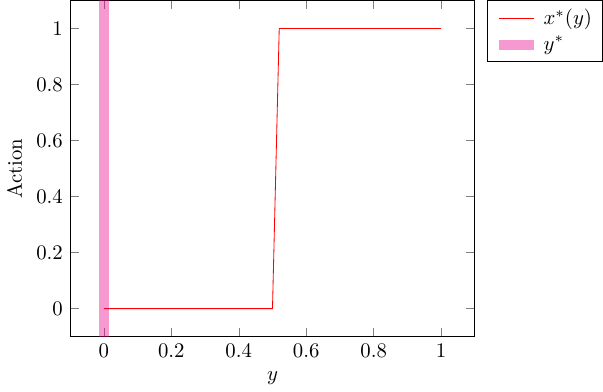}
        \caption{BR for $x$ as a function of $y$}
        \label{fig:fig:EagleGame_egalitarian x_br}
    \end{subfigure}
    \newline
    \begin{subfigure}[t]{0.45\textwidth}
        \centering
        \includegraphics[width=\textwidth]{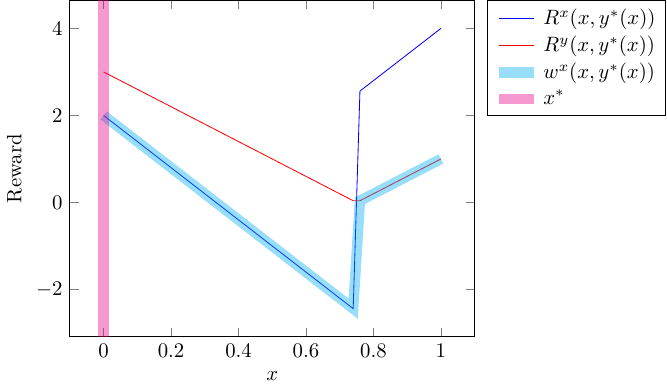}
        \caption{Rewards as a function of $x$}
        \label{fig:fig:EagleGame_egalitarian br_rewards_x}
    \end{subfigure}
    \hfill
    \begin{subfigure}[t]{0.45\textwidth}
        \centering
        \includegraphics[width=\textwidth]{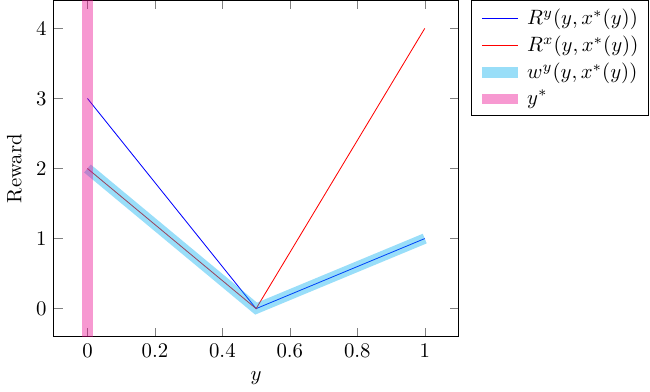}
        \caption{Rewards as a function of $y$}
        \label{fig:fig:EagleGame_egalitarian br_rewards_y}
    \end{subfigure}
    \caption{Egalitarian WE for EagleGame \\ $x^* = 0.000, y^* = 0.000, R^x = 2.000, R^y = 3.000$}
    \label{fig:EagleGame_egalitarian}
\end{figure*}

\end{document}